\RequirePackage{ifpdf}
\ifpdf 
\documentclass[pdftex]{sigma}
\else
\documentclass{sigma}
\fi

\usepackage{mathrsfs}

\def\tr{\mathrm{tr\,}}

\def\im{\mathrm{Im\,}}
\def\ad{\mathrm{ad\,}}

\usepackage{eucal}
\def\tr{\mathrm{tr\,}}

\def\im{\mathrm{Im\,}}
\def\ad{\mathrm{ad\,}}

\def\openone{\leavevmode\hbox{\small1\kern-3.3pt\normalsize1}}

\def\a{{\boldsymbol a}}
\def\b{{\boldsymbol b}}
\def\c{{\boldsymbol c}}
\def\d{{\boldsymbol d}}

\def\T{{\boldsymbol T}}
\def\S{{\boldsymbol S}}

\def\bPsi{{\boldsymbol \Psi}}
\def\bPhi{{\boldsymbol \Phi}}

\def\ad{\mbox{ad\,}}

\def\tr{\mbox{tr\,}}
\def\im{\mbox{Im\,}}
\def\diag{\mbox{diag\,}}

\def\otimescomma{\mathop{\otimes}\limits_{'}}
\def\wedgecomma{\mathop{\wedge}\limits_{'}}

\def\Res{\mathop{\mbox{Res}\,}\limits}

\def\openone{\leavevmode\hbox{\small1\kern-3.3pt\normalsize1}}

\allowdisplaybreaks

\usepackage{epsf}

\def\newpic#1{%
   \def\emline##1##2##3##4##5##6{%
      \put(##1,##2){\special{em:point #1##3}}%
      \put(##4,##5){\special{em:point #1##6}}%
      \special{em:line #1##3,#1##6}}}

\newpic{}

\def\wedgecomma{\mathop{\wedge}\limits_{'}}
\def\ad{\mbox{ad}\,}
\def\otimescomma{\mathop{\otimes}\limits_{'}}
\def\tr{\mbox{tr}\,}
\def\im{{\rm Im}\,}

\def\Aut{\mbox{Aut}\,}

\def\openone{\leavevmode\hbox{\small1\kern-3.3pt\normalsize1}}

\def\bbbz{{\Bbb Z}}
\newcommand{\rd}{\mathrm{d}} 
\newcommand{\re}{\mathrm{e}} 
\newcommand{\ri}{\mathrm{i}} 
\def\openone{\leavevmode\hbox{\small1\kern-3.3pt\normalsize1}}
\def\Res{\mathop{\mbox{Res}\,}\limits}

\def\ad{{\mbox{ad}}}
\def\Aut{{\mbox{Aut\,}}}

\allowdisplaybreaks
\def\tr{\mathrm{tr\,}}

\def\ad{\mathrm{ad\,}}

\usepackage{amssymb}
\usepackage{eucal}
\def\tr{\mathrm{tr\,}}

\def\ad{\mathrm{ad\,}}

\def\openone{\leavevmode\hbox{\small1\kern-3.3pt\normalsize1}}

\def\a{{\boldsymbol a}}
\def\b{{\boldsymbol b}}
\def\c{{\boldsymbol c}}
\def\d{{\boldsymbol d}}
\def\s{{\boldsymbol s}}

\def\T{{\boldsymbol T}}
\def\S{{\boldsymbol S}}

\def\0{{\boldsymbol 0}}
\def\bPsi{{\boldsymbol \Psi}}
\def\bPhi{{\boldsymbol \Phi}}

\def\bpsi{{\boldsymbol \psi}}
\def\bphi{{\boldsymbol \phi}}

\def\ad{\mbox{ad\,}}

\def\tr{\mbox{tr\,}}
\def\diag{\mbox{diag\,}}

\def\otimescomma{\mathop{\otimes}\limits_{'}}
\def\wedgecomma{\mathop{\wedge}\limits_{'}}

\def\bbbe{{\mathbb E}}
\def\bbbc{{\mathbb C}}
\def\bbbr{{\mathbb R}}

\begin{document}
\numberwithin{equation}{section}

\allowdisplaybreaks

\renewcommand{\PaperNumber}{029}

\FirstPageHeading

\renewcommand{\thefootnote}{$\star$}

\ShortArticleName{Reductions of Multicomponent mKdV Equations on
Symmetric Spaces of {\bf DIII}-Type}

\ArticleName{Reductions of Multicomponent mKdV Equations\\ on
Symmetric Spaces of {\bf DIII}-Type\footnote{This paper is a contribution to the Proceedings
of the Seventh International Conference ``Symmetry in Nonlinear
Mathematical Physics'' (June 24--30, 2007, Kyiv, Ukraine). The
full collection is available at
\href{http://www.emis.de/journals/SIGMA/symmetry2007.html}{http://www.emis.de/journals/SIGMA/symmetry2007.html}}}

\Author{Vladimir S. GERDJIKOV and Nikolay A. KOSTOV}

\AuthorNameForHeading{V.S. Gerdjikov and N.A. Kostov}

\Address{Institute for Nuclear Research and Nuclear
Energy, Bulgarian Academy of Sciences,\\
72 Tsarigradsko chaussee, 1784 Sof\/ia, Bulgaria}

\Email{\href{mailto:gerjikov@inrne.bas.bg}{gerjikov@inrne.bas.bg}, \href{mailto:nakostov@inrne.bas.bg}{nakostov@inrne.bas.bg}}

\ArticleDates{Received December 14, 2007, in f\/inal form February
27, 2008; Published online March 11, 2008}

\Abstract{New reductions for the multicomponent modif\/ied Korteveg--de Vries (MMKdV) equations on the symmetric spaces of {\bf
DIII}-type are derived using the approach based on the reduction
group introduced by A.V.~Mikhailov. The relevant inverse
scattering problem is studied and reduced to a Riemann--Hilbert
problem. The minimal sets of scattering data~$\mathcal{T}_i$,
$i=1,2$ which allow one to reconstruct uniquely both the
scattering matrix and the potential of the Lax operator are
def\/ined. The ef\/fect of the new reductions on the hierarchy of
Hamiltonian structures of MMKdV and on~$\mathcal{T}_i$ are
studied. We illustrate our results by the MMKdV equations related to
the algebra $\mathfrak{g}\simeq so(8)$ and derive several new
MMKdV-type equations using group of reductions isomorphic to
${\mathbb Z}_{2}$, ${\mathbb Z}_{3}$, ${\mathbb Z}_{4}$.}


\Classification{37K20; 35Q51; 74J30; 78A60}

\section{Introduction}

The modif\/ied Korteweg--de Vries equation \cite{Wad}
\begin{gather*}
q_t + q_{xxx} +6\epsilon q_xq^2(x,t)=0, \qquad \epsilon =\pm 1,
\end{gather*}
has natural multicomponent generalizations related to the
symmetric spaces \cite{AF*87}. They can be integrated by the ISM
using the fact that they allow the following Lax representation:
\begin{gather}\label{eq:LM}
L\psi \equiv  \left( i\frac{d}{dx} + Q(x,t) - \lambda J\right)
\psi(x,t,\lambda ) =0,   \\
Q(x,t) = \left(\begin{array}{cc} 0& q \\ p & 0 \\
\end{array}\right), \qquad J = \left(\begin{array}{cc} \openone& 0 \\ 0 & -\openone \\
\end{array}\right), \nonumber\\
M\psi \equiv  \left( i\frac{d}{dt} + V_0(x,t) + \lambda V_1(x,t)
+ \lambda^2 V_2(x,t) - 4 \lambda^3 J\right)\psi(x,t,\lambda )
 =\psi(x,t,\lambda )C(\lambda), \nonumber \\
V_2(x,t) = 4 Q(x,t), \qquad V_1(x,t) = 2i J Q_x + 2 J Q^2  , \qquad
V_0(x,t) = - Q_{xx} -2 Q^3,\nonumber
\end{gather}
where $J $ and $Q(x,t)$ are $2r\times 2r$ matrices: $ J$ is a
block diagonal and $Q(x,t)$ is a block-of\/f-diagonal matrix. The
corresponding MMKdV equations take the form
\begin{gather*}
\frac{\partial Q}{\partial t} +  \frac{\partial^3 Q}{\partial x^3}
+ 3 \left( Q_xQ^2 + Q^2Q_x\right) =0.
\end{gather*}

The analysis in \cite{AF*87,ForKu*83,APT} reveals a number of
important results. These include the corresponding multicomponent
generalizations of KdV equations and the generalized Miura
transformations interrelating them with the generalized MMKdV
equations; two of their most important reductions as well as their
Hamiltonians.

Our aim in this paper is to explore new types of reductions of the
MMKdV equations. To this end we make use of the reduction group
introduced by Mikhailov \cite{Mikh,LoMi1} which allows one to
impose algebraic reductions on the coef\/f\/icients of $Q(x,t)$ which
will be automatically compatible with the evolution of MMKdV.
Similar problems have been analyzed for the $N$-wave type
equations related to the simple Lie algebras of rank~2 and~3
\cite{vgn,vgrn} and the multicomponent NLS equations~\cite{manev04,Varna04}.
Here we illustrate our analysis by the MMKdV equations related to the
algebras $\mathfrak{g}\simeq so(2r)$ which are linked to the {\bf
DIII}-type symmetric spaces series. Due to the fact that the
dispersion law for MNLS is proportional to~$\lambda^2$ while for
MMKdV it is proportional to~$\lambda^3$ the sets of admissible
reductions for these two NLEE equations dif\/fer substantially.

\looseness=-1
In the next Section~\ref{sec2} we give some preliminaries on the scattering
theory for $L$, the reduction group and graded Lie algebras. In
Section~\ref{sec:FAS} we construct the fundamental analytic solutions of $L$,
formulate the corresponding Riemann--Hilbert problem and introduce
the minimal sets of scattering data $\mathcal{T}_i$, $i=1,2$ which
def\/ine uniquely both the scattering matrix and the solution of the
MMKdV $Q(x,t)$. Some of these facts have been discussed in more
details in \cite{Varna04}, others had to be modif\/ied and extended
so that they adequately take into account the peculiarities of the
{\bf DIII} type symmetric spaces. In particular we modif\/ied the
def\/inition of the fundamental analytic solution which lead to
changes in the formulation of the Riemann--Hilbert problem. In
Section~\ref{sec4} we show that the ISM can be interpreted as a generalized
Fourier \cite{Varna04} transform which maps the potential $Q(x,t)$
onto the minimal sets of scattering data $\mathcal{T}_i$. Here we
brief\/ly outline the hierarchy of Hamiltonian structures for the
generic MMKdV equations. The next Section~\ref{sec5} contains two classes
of nontrivial reductions of the MMKdV equations related to the algebra
$so(8)$. The f\/irst class is performed with automorphisms of
$so(8)$ that preserve $J$;  the second class uses automorphisms
that map $J$ into $-J$. While the reductions of f\/irst type can be
applied both to MNLS and MMKdV equations, the reductions of second
type can be applied only to MMKdV equations. Under them `half' of
the members of the Hamiltonian hierarchy become degenerated~\cite{DrSok,AF*87}. For both classes of reductions we f\/ind
examples with groups of reductions isomorphic to $\bbbz_2$,
$\bbbz_3$ and $\bbbz_4$. We also provide the corresponding reduced
Hamiltonians and symplectic forms and Poisson brackets. At the end
of Section~\ref{sec5} we derive the ef\/fects of these reductions on the
scattering matrix and on the minimal sets of scattering data. In
Section~\ref{ssec:5.3} following \cite{ForKu*83} we analyze  the classical
$r$-matrix for the corresponding NLEE. The ef\/fect of reductions on
the classical $r$-matrix is discussed. The last Section contains
some conclusions.

\section{Preliminaries}\label{sec2}
\subsection[Cartan-Weyl basis and Weyl group for $so(2r)$]{Cartan--Weyl basis and Weyl group for $\boldsymbol{so(2r)}$}\label{ssec:2.3}

Here we f\/ix the notations and the normalization conditions for the
Cartan--Weyl generators of $\mathfrak{g}\simeq so(2r) $, see e.g.~\cite{Helg}. The root system $\Delta $ of this series of simple
Lie algebras consists of the roots $\Delta \equiv \{ \pm
(e_i-e_j), \pm (e_i+e_j)\}$ where $1\leq i <j \leq r$. We
introduce an ordering in $\Delta$ by specifying the set of
positive roots $\Delta^+ \equiv \{ e_i-e_j, e_i+e_j\} $ for $1\leq
i <j \leq r$. Obviously all roots have the same length equal to 2.

We introduce the basis in the Cartan subalgebra by $h_k\in
\mathfrak{h} $, $k=1,\dots,r $ where $\{h_k\} $ are the Cartan
elements dual to the orthonormal basis $\{e_k\}$ in the root space
$\bbbe^r $. Along with $h_k $ we introduce also
\begin{gather*}
H_\alpha =  \sum_{k=1}^{r} (\alpha ,e_k) h_k, \qquad \alpha \in
\Delta ,
\end{gather*}
where $(\alpha ,e_k) $ is the scalar product in the root space
$\bbbe^r $ between the root $\alpha $ and $e_k $. The basis in
$so(2r)$ is completed by adding the Weyl generators $E_\alpha $,
$\alpha \in \Delta $.

The commutation relations for the elements of the Cartan--Weyl
basis are given by \cite{Helg}
\begin{gather*}
 [h_k,E_\alpha ] = (\alpha ,e_k) E_\alpha , \qquad [E_\alpha
,E_{-\alpha }]=H_\alpha , \\
 [E_\alpha ,E_\beta ] = \left\{ \begin{array}{ll} N_{\alpha
,\beta } E_{\alpha +\beta } \quad & \mbox{for}\; \alpha +\beta \in
\Delta, \\ 0 & \mbox{for}\; \alpha +\beta \not\in \Delta \cup\{0\}.
\end{array} \right.
\end{gather*}
We will need also the typical $2r$-dimensional representation of
$so(2r)$. For convenience we choose the following def\/inition for
the orthogonal algebras and groups
\begin{gather}\label{eq:so}
X\in so(2r) \longrightarrow X +S_0X^T \hat{S}_0 =0, \qquad T\in
SO(2r) \longrightarrow S_0T^T \hat{S}_0 =\hat{T},
\end{gather}
where by `hat' we denote the inverse matrix $\hat{T}\equiv T^{-1}$
and
\begin{gather}\label{eq:s0}
S_0 \equiv \sum_{k=1}^r (-1)^{k+1} \left( E_{k,\bar{k}} +
E_{\bar{k},k} \right) = \left(\begin{array}{cc}  0 & \s_0 \\
\hat{\s}_0 & 0 \\ \end{array} \right), \qquad \bar{k}=2r+1-k.
\end{gather}
Here and below by $E_{jk}$ we denote a $2r \times 2r$ matrix with
just one non-vanishing and equal to 1 matrix element at $j,k$-th
position: $(E_{jk})_{mn}=\delta_{jm}\delta_{kn}$. Obviously
$S_0^2=\openone$.
 In order to have the Cartan generators represented by
diagonal matrices we modif\/ied the def\/inition of orthogonal matrix,
see (\ref{eq:so}). Using the matrices $E_{jk}$ def\/ined by equation~(\ref{eq:s0}) we get
\begin{gather*}
h_k = E_{kk}-E_{\bar{k}\bar{k}}, \qquad E_{e_i-e_j} = E_{ij} -
(-1)^{i+j} E_{\bar{j}\bar{i}},  \\
E_{e_i+e_j} = E_{i\bar{j}} - (-1)^{i+j} E_{\bar{j}\bar{i}},
\qquad E_{-\alpha}=E_\alpha^T,
\end{gather*}
where $\bar{k}=2r+1-k$.

We will denote by $\vec{a}=\sum\limits_{k=1}^{r} e_k $ the $r
$-dimensional vector dual to $J\in \mathfrak{h} $; obviously
$J=\sum\limits_{k=1}^{r}h_k $. If the root $\alpha \in\Delta _+ $ is
positive (negative) then $(\alpha ,\vec{a})\geq 0 $ ($(\alpha
,\vec{a})<0 $ respectively). The normali\-zation of the basis is
determined by
\begin{gather*}
E_{-\alpha } =E_\alpha ^T, \qquad \langle E_{-\alpha },E_\alpha
\rangle =2,  \qquad  N_{-\alpha ,-\beta } = -N_{\alpha ,\beta }.
\end{gather*}
The root system $\Delta $ of $\mathfrak{g} $ is invariant with
respect to the Weyl ref\/lections $S_\alpha $; on the vectors
$\vec{y}\in \bbbe^r $ they act as
\begin{gather*}
S_\alpha \vec{y} = \vec{y} - {2(\alpha ,\vec{y}) \over (\alpha
,\alpha )} \alpha , \qquad \alpha \in \Delta .
\end{gather*}
All Weyl ref\/lections $S_\alpha $ form a f\/inite group
$W_{\mathfrak{g}} $ known as the Weyl group. On the root space
this group is isomorphic to $\mathcal{S}_r\otimes (\bbbz_2)^{r-1}$
where $\mathcal{S}_r$ is the group of permutations of the basic
vectors $e_j \in \bbbe^r$. Each of the $\bbbz_2$ groups acts on
$\bbbe^r$ by changing simultaneously the signs of two of the basic
vectors $e_j$.

One may introduce also an action of the Weyl group on the
Cartan--Weyl basis, namely \cite{Helg}
\begin{gather*}
 S_\alpha (H_\beta ) \equiv A_\alpha H_\beta A^{-1}_{\alpha } =
H_{S_\alpha \beta }, \nonumber\\
 S_\alpha (E_\beta ) \equiv A_\alpha E_\beta A^{-1}_{\alpha } =
n_{\alpha ,\beta } E_{S_\alpha \beta }, \qquad n_{\alpha ,\beta
}=\pm 1.
\end{gather*}

The matrices $A_\alpha $ are given (up to a factor from the Cartan
subgroup) by
\begin{gather}\label{eq:32.4}
A_\alpha =e^{E_\alpha } e^{-E_{-\alpha }} e^{E_\alpha } H_A,
\end{gather}
where $H_A $ is a conveniently chosen element from the Cartan
subgroup such that $H_A^2=\openone $. The formula (\ref{eq:32.4})
and the explicit form of the Cartan--Weyl basis in the typical
representation will be used in calculating the reduction condition
following from (\ref{eq:2.1}).

\subsection{Graded Lie algebras}\label{ssec:2.4}

One of the important notions in constructing integrable equations
and their reductions is the one of graded Lie algebra and
Kac--Moody algebras \cite{Helg}. The standard construction is based
on a~f\/inite order automorphism $C\in \Aut \mathfrak{g} $,
$C^N=\openone $. The eigenvalues of $C $ are $\omega ^k $,
$k=0,1,\dots , N-1 $, where $\omega =\exp(2\pi i/N) $. To each
eigenvalue there corresponds a linear subspace $\mathfrak{g}^{(k)}
\subset \mathfrak{g}$ determined by
\begin{gather*}
\mathfrak{g}^{(k)} \equiv \big\{ X\colon X\in \mathfrak{g}, \
C(X)=\omega ^k X \big\} .
\end{gather*}
Then $\mathfrak{g}=\mathop{\oplus}\limits_{k=0}^{N-1}
\mathfrak{g}^{(k)} $ and the grading condition holds
\begin{gather}\label{eq:34.1}
\left[\mathfrak{g}^{(k)} , \mathfrak{g}^{(n)} \right] \subset
\mathfrak{g}^{(k+n)},
\end{gather}
where $k+n $ is taken modulo $N $. Thus to each pair
$\{\mathfrak{g} , C\} $ one can relate an inf\/inite-dimensional
algebra of Kac--Moody type $\widehat{\mathfrak{g}}_C $ whose
elements are
\begin{gather}\label{eq:34.2}
X(\lambda ) = \sum_{k}^{} X_k \lambda ^k, \qquad X_k \in
\mathfrak{g}^{(k)} .
\end{gather}
The series in (\ref{eq:34.2}) must contain only f\/inite number of
negative (positive) powers of $\lambda $ and $\mathfrak{g}^{(k+N)}
\equiv \mathfrak{g}^{(k)}$. This construction is a most natural
one for Lax pairs; we see that due to the grading condition
(\ref{eq:34.1}) we can always impose a reduction on $L(\lambda ) $
and $M(\lambda ) $ such that both $U(x,t,\lambda ) $ and
$V(x,t,\lambda )\in \widehat{\mathfrak{g}}_C $. In the case of
symmetric spaces $N=2$ and $C$ is the Cartan involution. Then one
can choose the Lax operator $L$ in such a way that
\begin{gather*}
Q \in \mathfrak{g}^{(1)},  \qquad J \in \mathfrak{g}^{(0)}
\end{gather*}
as it is the case in (\ref{eq:LM}). Here the subalgebra
$\mathfrak{g}^{(0)}$ consists of all elements of $\mathfrak{g}$
commuting with~$J$. The special choice of $J=\sum\limits_{k=1}^r h_k$
taken above allows us to split the set of all positive roots
$\Delta^+$ into two subsets
\begin{gather*}
\Delta^+ =\Delta^+_0\cup \Delta^+_1, \qquad \Delta^+_0 = \{
e_i-e_j\}_{i<j}, \qquad  \Delta^+_1 = \{ e_i+e_j\}_{i<j}.
\end{gather*}
Obviously the elements $\alpha \in \Delta^+_1$ have the property
$\alpha(J) = (\alpha, \vec{a}) =2$, while the elements $\beta \in
\Delta^+_0$ have the property $\beta(J) = (\beta, \vec{a}) =0$.
In this section we outline some of the well known facts about the
spectral theory of the Lax operators of the type~(\ref{eq:LM}).

\subsection[The scattering problem for $L$]{The scattering problem for $\boldsymbol{L}$}\label{ssec:2.1}

Here we brief\/ly outline the  basic facts about the direct and the
inverse scattering problems
\cite{ZaSha,Sh,Sha,ZMNP,CaDe,CaDeB,FaTa,CaDe2,D1,BeSat,LMP} for
the system (\ref{eq:LM}) for the class of potentials $Q(x,t) $
that are smooth enough and fall of\/f to zero fast enough for
$x\to\pm\infty $ for all $t $. In what follows we  treat {\bf
DIII}-type symmetric spaces which means that $Q(x,t)$ is an
element of the algebra~$so(2r)$. In the examples below  we take
$r=4$ and $\mathfrak{g}\simeq so(8)$.

The main tool for solving the direct and inverse scattering
problems are the Jost solutions which are fundamental solutions
def\/ined by their asymptotics at $x\to\pm\infty $
\begin{gather*}
\lim_{x\to\infty } \psi (x,\lambda )\re^{\ri\lambda Jx} =\openone
, \qquad \lim_{x\to -\infty } \phi (x,\lambda )\re^{\ri\lambda Jx}
=\openone.
\end{gather*}

Along with the Jost solutions we introduce
\begin{gather*}
\xi(x,\lambda ) =\psi (x,\lambda )\re^{\ri\lambda Jx}, \qquad
\varphi (x,\lambda ) =\phi (x,\lambda )\re^{\ri\lambda Jx},
\end{gather*}
which satisfy the following linear integral equations
\begin{gather}\label{eq:5.2}
\xi(x,\lambda ) = \openone + \ri \int_{\infty }^{x} \rd y
\re^{-\ri\lambda J(x-y)} Q(y) \xi(y,\lambda ) \re^{\ri\lambda
J(x-y)},
\\
\label{eq:5.2'} \varphi (x,\lambda ) = \openone + \ri
\int_{-\infty }^{x} \rd y \re^{-\ri\lambda J(x-y)} Q(y) \varphi
(y,\lambda ) \re^{\ri\lambda J(x-y)}.
\end{gather}

These are Volterra type equations which, have solutions providing
one can ensure the convergence of the integrals in the right hand
side. For $\lambda  $ real the exponential factors in
(\ref{eq:5.2}) and (\ref{eq:5.2'}) are just oscillating and the
convergence is ensured by the fact that $Q(x,t)$ is quickly
vanishing for $x\to\infty$.

\begin{remark}\label{rem:n1}
It is an well known fact that if the potential $Q(x,t)\in so(2r)$
then the corresponding Jost solutions of equation~(\ref{eq:LM}) take
values in the corresponding group, i.e.~$\psi(x,\lambda),
\phi(x,\lambda)\in SO(2r)$.
\end{remark}

The Jost solutions as whole can not be extended for $\im \lambda
\neq 0 $. However some of their columns can be extended for
$\lambda \in \bbbc_+ $, others -- for $\lambda \in \bbbc_- $. More
precisely we can write down the Jost solutions $\psi (x,\lambda )
$ and $\phi (x,\lambda ) $ in the following block-matrix form
\begin{gather*}
\psi (x,\lambda ) = \left(|\psi^- (x,\lambda )\rangle , |\psi^+
(x,\lambda )\rangle \right), \qquad  \phi (x,\lambda ) =
\left(|\phi^+ (x,\lambda ) \rangle , |\phi^- (x,\lambda )\rangle
\right), \nonumber\\
|\psi^\pm (x,\lambda )\rangle = \left( \begin{array}{c}
  \bpsi_1^\pm (x,\lambda ) \\ \bpsi_2^\pm (x,\lambda ) \\
\end{array}\right), \qquad |\phi^\pm (x,\lambda )\rangle
= \left( \begin{array}{c} \bphi_1^\pm (x,\lambda ) \\
\bphi_2^\pm (x,\lambda ) \\ \end{array}\right),
\end{gather*}
where the superscript $+ $ and (resp.~$- $) shows that the
corresponding $r \times r$  block-matrices allow analytic
extension for $\lambda \in \bbbc_+ $ (resp.~$\lambda \in \bbbc_- $).

Solving the direct scattering problem means given the potential
$Q(x) $ to f\/ind the scattering matrix $T(\lambda ) $. By
def\/inition $T(\lambda ) $ relates the two Jost solutions
\begin{gather}\label{eq:6.1}
\phi (x,\lambda ) =\psi (x,\lambda )T(\lambda ), \qquad T(\lambda
)= \left(\begin{array}{cc} \a^+(\lambda ) & -\b^-(\lambda )
\\
\b^+(\lambda ) & \a^-(\lambda ) \end{array}\right)
\end{gather}
and has compatible block-matrix structure.  In what follows we
will need also the inverse of the scattering matrix
\begin{gather*}
\psi (x,\lambda ) =\phi (x,\lambda )\hat{T}(\lambda ), \qquad
\hat{T}(\lambda )\equiv \left(\begin{array}{cc} \c^-(\lambda ) &
\d^-(\lambda ) \\ -\d^+(\lambda ) & \c^+(\lambda )
\end{array}\right),
\end{gather*}
where
\begin{gather}
\c^-(\lambda ) = \hat{\a}^+(\lambda ) (\openone +\rho ^-\rho
^+)^{-1} = (\openone +\tau^+\tau^-)^{-1} \hat{\a}^+(\lambda ), \nonumber\\
\d^-(\lambda ) = \hat{\a}^+(\lambda )\rho ^-
(\lambda ) (\openone  +\rho ^+\rho ^-)^{-1} = (\openone +\tau^+
\tau^-)^{-1} \tau^+ (\lambda ) \hat{\a}^-(\lambda ), \nonumber\\
\c^+(\lambda ) = \hat{\a}^-(\lambda )
(\openone +\rho^+\rho ^-)^{-1}= (\openone +\tau^-
\tau^+)^{-1}\hat{\a}^-(\lambda ), \nonumber\\
\d^+(\lambda ) = \hat{\a}^-(\lambda )\rho
^+(\lambda ) (\openone  +\rho ^-\rho ^+)^{-1}= (\openone
+\tau^-\tau^+)^{-1} \tau^- (\lambda )\hat{\a}^+(\lambda ).\label{eq:6.2}
\end{gather}
The diagonal blocks of  $T(\lambda ) $ and $\hat{T}(\lambda ) $
allow analytic continuation of\/f the real axis, namely
$\a^+(\lambda ) $, $\c^+(\lambda ) $ are analytic functions of
$\lambda  $ for $\lambda \in \bbbc_+ $, while  $\a^-(\lambda ) $,
$\c^-(\lambda ) $  are analytic functions of $\lambda$ for
$\lambda \in \bbbc_- $. We introduced also $\rho ^\pm(\lambda ) $
and $\tau^\pm(\lambda ) $ the multicomponent generalizations of
the ref\/lection coef\/f\/icients (for the scalar case, see
\cite{AKNS,CaDe,KN79})
\begin{gather*}
\rho ^\pm(\lambda ) =\b^\pm\hat{\a}^\pm (\lambda ) =\hat{\c}^\pm
\d^\pm(\lambda ), \qquad \tau^\pm(\lambda ) =\hat{\a}^\pm
\b^\mp(\lambda ) =\d^\mp\hat{\c}^\pm (\lambda ).
\end{gather*}
The ref\/lection coef\/f\/icients do not have analyticity properties and
are def\/ined only for $\lambda \in \bbbr$.

From Remark~\ref{rem:n1} one concludes that $T(\lambda)\in
SO(2r)$, therefore it must satisfy the second of the equations in
(\ref{eq:so}). As a result we get the following relations between
$\c^\pm$, $\d^\pm$ and $\a^\pm$, $\b^\pm$
\begin{gather}
\c^+(\lambda) = \hat{\s_0} \a^{+,T}(\lambda)  \s_0, \qquad
\c^-(\lambda) = \s_0\a^{-,T}(\lambda)  \hat{\s}_0, \nonumber\\
\d^+(\lambda) = -\hat{\s_0} \b^{+,T}(\lambda)  \s_0, \qquad
\d^-(\lambda) = -\s_0\b^{-,T}(\lambda)  \hat{\s}_0,\label{eq:cd-ab}
\end{gather}
and in addition we have
\begin{gather}
\rho^+(\lambda) = -\hat{\s}_0 \rho^{+,T}(\lambda)  \s_0, \qquad
\rho^-(\lambda) = -\s_0\rho^{-,T}(\lambda)  \hat{\s}_0, \nonumber\\
\tau^+(\lambda) =  -\s_0\tau^{+,T}(\lambda)  \hat{\s}_0, \qquad
\tau^-(\lambda) = -\hat{\s}_0 \tau^{-,T}(\lambda)  \s_0.\label{eq:ro-roT}
\end{gather}

Next we need also the asymptotics of the Jost solutions and the
scattering matrix for $\lambda \to\infty $
\begin{gather*}
 \lim_{\lambda \to -\infty } \phi (x,\lambda ) \re^{\ri\lambda
Jx} = \lim_{\lambda \to\infty } \psi (x,\lambda ) \re^{\ri\lambda
Jx} =\openone ,
\qquad \lim_{\lambda \to\infty } T(\lambda )  =\openone ,\\
 \lim_{\lambda \to\infty } \a^+(\lambda )  = \lim_{\lambda
\to\infty } \c^-(\lambda )  = \openone , \qquad \lim_{\lambda
\to\infty } \a^-(\lambda )  = \lim_{\lambda \to\infty }
\c^+(\lambda )  = \openone . \nonumber
\end{gather*}

The inverse to the Jost solutions $\hat{\psi }(x,\lambda) $ and
$\hat{\phi }(x,\lambda) $ are solutions to
\begin{gather}\label{eq:L-inv}
\ri {\rd\hat{\psi }\over \rd x } - \hat{\psi }(x,\lambda)
(Q(x)-\lambda J) =0,
\end{gather}
satisfying the conditions
\begin{gather}\label{eq:J-inv}
\lim_{x\to\infty } \re^{-\ri\lambda Jx}\hat{\psi
}(x,\lambda)=\openone , \qquad \lim_{x\to -\infty }
\re^{-\ri\lambda Jx}\hat{\phi }(x,\lambda)=\openone .
\end{gather}

Now it is the collections of rows of $\hat{\psi }(x,\lambda) $ and
$\hat{\phi }(x,\lambda) $ that possess analytic properties in
$\lambda  $
\begin{gather}
\hat{\psi }(x,\lambda) = \left( \begin{array}{c} \langle
\hat{\psi }^+(x,\lambda) | \\ \langle \hat{\psi }^-(x,\lambda) |
\end{array} \right), \qquad
\hat{\phi }(x,\lambda) = \left( \begin{array}{c} \langle
\hat{\bphi }^-(x,\lambda) | \\ \langle \hat{\bphi }^+(x,\lambda) |
\end{array} \right),\nonumber\\
\langle \hat{\psi }^\pm(x,\lambda) |  = \left( \s_0^{\pm 1}
\bpsi_2^\pm, \s_0^{\pm 1} \bpsi_1^\pm \right) (x,\lambda), \qquad
\langle \hat{\phi }^\pm(x,\lambda) | = \left( \s_0^{\mp 1}
\bpsi_2^\pm, \s_0^{\mp 1} \bpsi_1^\pm \right) (x,\lambda).\label{eq:psi-inv}
\end{gather}
Just like the Jost solutions, their inverse (\ref{eq:psi-inv}) are
solutions to linear equations (\ref{eq:L-inv}) with regular
boundary conditions (\ref{eq:J-inv}); therefore they can have no
singularities on the  real axis $\lambda \in\bbbr$. The same holds
true also for the scattering matrix $T(\lambda )=\hat{\psi
}(x,\lambda) \phi (x,\lambda) $ and its inverse $\hat{T}(\lambda
)=\hat{\phi }(x,\lambda) \psi (x,\lambda) $, i.e.
\begin{gather*}
\a^+(\lambda ) = \langle \hat{\psi }^+(x,\lambda) |\phi^+
(x,\lambda) \rangle , \qquad \a^-(\lambda ) = \langle \hat{\psi
}^-(x,\lambda) |\phi^- (x,\lambda) \rangle ,
\end{gather*}
as well as
\begin{gather*}
\c^+(\lambda ) = \langle \hat{\phi }^+(x,\lambda) |\psi^+
(x,\lambda) \rangle , \qquad \c^-(\lambda ) = \langle \hat{\phi
}^-(x,\lambda) |\psi^- (x,\lambda) \rangle ,
\end{gather*}
are analytic for $\lambda \in \bbbc_\pm $ and have no
singularities for $\lambda \in \bbbr$.  However they may become
degenerate (i.e., their determinants may vanish) for some values
$\lambda _j^\pm \in \bbbc_\pm $ of $\lambda  $. Below we brief\/ly
analyze the structure of these degeneracies and show that they are
related to discrete spectrum of~$L$.

\section[The fundamental analytic solutions and the
Riemann-Hilbert problem]{The fundamental analytic solutions\\ and the
Riemann--Hilbert problem}\label{sec:FAS}

\subsection{The fundamental analytic solutions }\label{ssec:FAS}

The next step is to construct the fundamental analytic solutions
(FAS) $\chi ^\pm(x,\lambda )$ of (\ref{eq:LM}). Here we slightly
modify the def\/inition in \cite{Varna04} to ensure that $\chi
^\pm(x,\lambda )\in SO(2r)$. Thus we def\/ine
\begin{gather}
\chi ^+(x,\lambda ) \equiv  \left(|\phi ^+\rangle , |\psi
^+\hat{c}^+\rangle\right)(x,\lambda) = \phi (x,\lambda )
\S^+(\lambda ) = \psi (x,\lambda ) \T^-(\lambda )D^+(\lambda ) ,\nonumber
\\ \chi ^-(x,\lambda ) \equiv  \left(|\psi ^-\hat{c}^-\rangle ,
|\phi ^-\rangle \right)(x,\lambda ) = \phi (x,\lambda )
\S^-(\lambda )=\psi (x,\lambda )\T^+(\lambda ) D^-(\lambda ) ,\label{eq:6.3}
\end{gather}
where the block-triangular functions $\S^\pm(\lambda ) $ and
$\T^\pm(\lambda ) $ are given by
\begin{gather}
\S^+(\lambda ) = \left( \begin{array}{cc} \openone  &
\d^-\hat{\c}^+(\lambda )\\ 0 & \openone \end{array}\right),\qquad
\T^-(\lambda ) =  \left( \begin{array}{cc} \openone & 0 \\
\b^+\hat{\a}^+(\lambda ) & \openone  \end{array}\right), \nonumber\\
\S^-(\lambda ) =  \left( \begin{array}{cc} \openone & 0 \\
-\d^+\hat{\c}^-(\lambda ) & \openone  \end{array}\right), \qquad
\T^+(\lambda ) = \left( \begin{array}{cc} \openone  &
-\b^-\hat{\a}^-(\lambda ) \\ 0 & \openone \end{array}\right).\label{eq:6.4}
\end{gather}
The matrices $D^\pm(\lambda )$ are  block-diagonal and equal
\begin{gather*}
D^+(\lambda ) = \left( \begin{array}{cc} \a^+(\lambda ) &0 \\ 0 &
\hat{\c}^+(\lambda ) \end{array}\right), \qquad D^-(\lambda ) =
\left(\begin{array}{cc} \hat{\c}^-(\lambda ) &0 \\ 0 &
\a^-(\lambda )
\end{array}\right).
\end{gather*}
The upper scripts $\pm$ here refer to their analyticity properties
for $\lambda \in\bbbc_\pm$.

In view of the relations (\ref{eq:cd-ab}) it is easy to check that
all factors $\S^\pm $, $\T^\pm $ and $D^\pm $ take values in the
group $SO(2r)$. Besides, since
\begin{gather}
T(\lambda ) = \T^-(\lambda )D^+(\lambda )\hat{\S}^+(\lambda ) =
\T^+(\lambda )D^-(\lambda )\hat{\S}^-(\lambda ), \nonumber\\
\hat{T}(\lambda ) = \S^+(\lambda )\hat{D}^+(\lambda )
\hat{\T}^-(\lambda ) = \S^-(\lambda )\hat{D}^-(\lambda )
\hat{\T}^+(\lambda ), \label{eq:7.1}
\end{gather}
we can view the factors $\S^\pm $, $\T^\pm $ and $D^\pm $ as
generalized Gauss decompositions (see \cite{Helg}) of $T(\lambda )
$ and its inverse.

The relations between $\c^\pm(\lambda ) $, $\d^\pm(\lambda ) $ and
$\a^\pm(\lambda ) $, $\b^\pm(\lambda ) $ in equation (\ref{eq:6.2})
ensure that equations~(\ref{eq:7.1}) become identities. From equations
(\ref{eq:6.3}), (\ref{eq:6.4}) we derive
\begin{gather}\label{eq:9.4}
\chi ^+(x,\lambda ) = \chi ^-(x,\lambda ) G_0(\lambda ), \qquad
 \chi ^-(x,\lambda ) = \chi ^+(x,\lambda )
\hat{G}_0(\lambda ), \\
\label{eq:11.3} G_0(\lambda ) = \left(\begin{array}{cc} \openone
& \tau^+ \\ \tau^- & \openone + \tau^-\tau^+  \end{array}
\right), \qquad \hat{G}_0(\lambda ) = \left(\begin{array}{cc}
\openone + \tau^+\tau^- & -\tau^+ \\ -\tau^- & \openone
 \end{array} \right)
\end{gather}
valid for $\lambda \in \bbbr$. Below  we introduce
\begin{gather*}
X^\pm(x,\lambda ) = \chi ^\pm(x,\lambda ) \re^{\ri\lambda Jx}.
\end{gather*}
Strictly speaking it is $X^\pm(x,\lambda)$ that allow analytic
extension for $\lambda \in \bbbc_\pm$. They have also another nice
property, namely their asymptotic behavior for $\lambda
\to\pm\infty  $ is given by
\begin{gather}\label{eq:12.1}
\lim_{\lambda \to\infty } X^\pm(x,\lambda ) =\openone .
\end{gather}
Along with $X^\pm(x,\lambda)$ we can use another set of FAS
$\tilde{X}^\pm(x,\lambda)= X^\pm(x,\lambda)\hat{D}^\pm$, which
also satisfy equation (\ref{eq:12.1}) due to the fact that
\begin{gather*}
\lim_{\lambda \to\infty } D^\pm(\lambda ) =\openone.
\end{gather*}
The analyticity properties of $X^\pm(x,\lambda ) $  and
$\tilde{X}^\pm(x,\lambda ) $ for $\lambda \in\bbbc_\pm $ along
with equation~(\ref{eq:12.1}) are crucial for our considerations.

\subsection[The Riemann-Hilbert problem]{The Riemann--Hilbert problem}\label{ssec:RHP}

The equations (\ref{eq:9.4}) and (\ref{eq:11.3}) can be written down
as
\begin{gather}\label{eq:12.4}
X^+(x,\lambda ) = X^-(x,\lambda ) G(x,\lambda ), \qquad \lambda
\in \bbbr,
\end{gather}
where
\begin{gather*}
G(x,\lambda ) = \re^{-\ri\lambda Jx} G_0(\lambda )\re^{\ri\lambda
Jx}.
\end{gather*}

Likewise the second pair of FAS satisfy
\begin{gather}\label{eq:12.5}
\tilde{X}^+(x,\lambda ) = \tilde{X}^-(x,\lambda )
\tilde{G}(x,\lambda ), \qquad \lambda \in \bbbr
\end{gather}
with
\begin{gather*}
\tilde{G}(x,\lambda ) = \re^{-\ri\lambda Jx} \tilde{G}_0(\lambda
)\re^{\ri\lambda Jx} \qquad  \tilde{G}_0(\lambda ) =
\left(\begin{array}{cc} \openone +\rho^-\rho^+ & \rho^- \\
\rho^+ & \openone \\ \end{array} \right).
\end{gather*}

Equation (\ref{eq:12.4}) (resp.\ equation (\ref{eq:12.5})) combined with
(\ref{eq:12.1}) is known in the literature \cite{Gakhov} as a
Riemann--Hilbert problem (RHP) with canonical normalization. It is
well known that RHP with canonical normalization has unique
regular solution; the matrix-valued solutions $X_0^+(x,\lambda ) $
and $X_0^-(x,\lambda ) $ of (\ref{eq:12.4}), (\ref{eq:12.1}) is
called regular if $\det X_0^\pm(x,\lambda ) $ does not vanish for
any $\lambda \in\bbbc_\pm $.

\begin{figure}[t]
\centerline{\includegraphics*{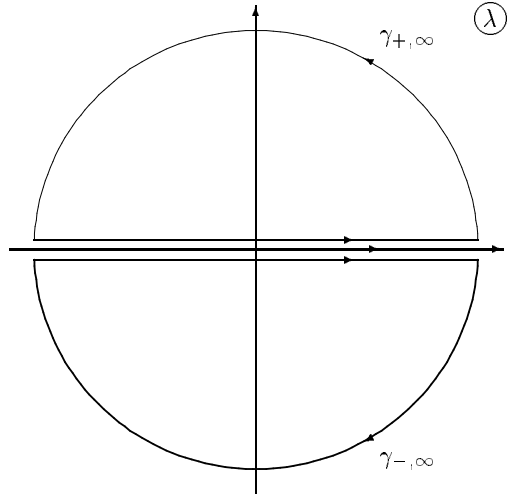}}
\caption{The contours $\gamma _\pm =\bbbr\cup\gamma_{\pm\infty
}$.} \label{fig:1}
\end{figure}

Let us now apply the contour-integration method to derive the
integral decompositions of~$X^\pm(x,\lambda ) $. To this end we
consider the contour integrals
\begin{gather*}
\mathcal{J}_1(\lambda ) = {1 \over 2\pi \ri} \oint_{\gamma _+}
{\rd\mu \over \mu -\lambda } X^+(x,\mu ) - {1 \over 2\pi \ri}
\oint_{\gamma _-} {\rd\mu \over \mu -\lambda } X^-(x,\mu ),
\end{gather*}
and
\begin{gather*}
\mathcal{J}_2(\lambda ) = {1 \over 2\pi \ri} \oint_{\gamma _+}
{\rd\mu \over \mu -\lambda} \tilde{X}^+(x,\mu)- {1 \over 2\pi \ri}
\oint_{\gamma _-}{\rd\mu \over\mu -\lambda }\tilde{X}^-(x,\mu ),
\end{gather*}
where $\lambda \in\bbbc_+ $ and the contours $\gamma _\pm $ are
shown on Fig.~\ref{fig:1}.

Each of these integrals can be evaluated by Cauchy residue
theorem. The result for $\lambda\in \bbbc_+$ are
\begin{gather}\label{eq:J_11}
\mathcal{J}_1(\lambda ) = X^+(x,\lambda ) + \sum_{j =1}^{N}
\Res_{\mu =\lambda _j^+} {X^+(x,\mu ) \over \mu -\lambda  } +
\sum_{j =1}^{N} \Res_{\mu =\lambda _j^-} {X^-(x,\mu ) \over \mu
-\lambda  }, \\
\label{eq:J_11t} \mathcal{J}_2(\lambda ) = \tilde{X}^+(x,\lambda )
+ \sum_{j =1}^{N} \Res_{\mu =\lambda _j^+} {\tilde{X}^+(x,\mu )
\over \mu -\lambda  } + \sum_{j =1}^{N} \Res_{\mu =\lambda _j^-}
{\tilde{X}^-(x,\mu ) \over \mu -\lambda  }.
\end{gather}
The discrete sums in the right hand sides of equations (\ref{eq:J_11})
and (\ref{eq:J_11t})  naturally provide the contribution from the
discrete spectrum of $L$.  For the sake of simplicity we assume
that $L $ has a f\/inite number of simple eigenvalues $\lambda
_j^\pm\in \bbbc_\pm $; for additional details see \cite{Varna04}.
Let us clarify the above statement. For the $2\times 2$
Zakharov--Shabat problem it is well known that the discrete
eigenvalues of $L$ are provided by the zeroes of the transmission
coef\/f\/icients $a^\pm(\lambda)$, which in that case are scalar
functions. For the more general $2r\times 2r$ Zakharov--Shabat
system~(\ref{eq:LM}) the situation becomes more complex because
now $a^\pm(\lambda)$ are $r\times r$ matrices. The discrete
eigenvalues~$\lambda_j^\pm$ now are the points at which
$a^\pm(\lambda)$ become degenerate and their inverse develop pole
singularities. More precisely, we assume that in the vicinities of
$\lambda_j^\pm$ $\a^\pm(\lambda)$, $\c^\pm(\lambda)$ and their
inverse  $\hat{\a}^\pm(\lambda)$, $\hat{\c}^\pm(\lambda)$ have the
following decompositions in Laurent series
\begin{gather*}
 \a^\pm(\lambda) = \a_j^\pm + (\lambda
-\lambda_j^\pm) \dot{\a}_j^\pm + \cdots, \qquad \c^\pm(\lambda)
= \c_j^\pm + (\lambda -\lambda_j^\pm) \dot{\c}_j^\pm + \cdots,
\\ 
 \hat{\a}^\pm(\lambda) =
\frac{\hat{\a}_j^\pm}{\lambda -\lambda_j^\pm} +
\hat{\dot{\a}}_j^\pm + \cdots, \qquad \hat{\c}^\pm(\lambda) =
\frac{\hat{\c}_j^\pm}{\lambda -\lambda_j^\pm} +
\hat{\dot{\a}}_j^\pm + \cdots,
\end{gather*}
where all the leading coef\/f\/icients $ \a_j^\pm$, $\hat{\a}_j^\pm$ $
\c_j^\pm$, $\hat{\c}_j^\pm$ are degenerate matrices such that
\begin{gather*}
 \hat{\a}_j^\pm \a_j^\pm = \a_j^\pm \hat{\a}_j^\pm =0, \qquad
  \hat{\c}_j^\pm \c_j^\pm =  \c_j^\pm \hat{\c}_j^\pm =0.
\end{gather*}
In addition we have more relations such as
\begin{gather*}
\hat{\a}_j^\pm \dot{\a}_j^\pm + \hat{\dot{\a}}_j^\pm \a_j^\pm
=\openone, \qquad \hat{\c}_j^\pm \dot{\c}_j^\pm +
\hat{\dot{\c}}_j^\pm \c_j^\pm =\openone,
\end{gather*}
that are needed to ensure that the identities
$\hat{\a}^\pm(\lambda) \a^\pm(\lambda) =\openone $,
$\hat{\c}^\pm(\lambda) \c^\pm(\lambda) =\openone $ etc hold true
for all values of $\lambda$.

The assumption that the eigenvalues are simple here means that we
have considered only f\/irst order pole singularities of
$\hat{\a}_j^\pm(\lambda)$ and $\hat{\c}_j^\pm(\lambda)$. After
some additional considerations we f\/ind that the `halfs' of the
Jost solutions $|\psi^\pm (x,\lambda)\rangle$ and $|\phi^\pm
(x,\lambda) \rangle$ satisfy the following relationships for
$\lambda =\lambda_j^\pm$
\begin{gather*}
|\psi_j^\pm (x)\hat{\c}_j^\pm \rangle = \pm |\phi_j^\pm (x)
\tau_j^\pm \rangle, \qquad |\phi_j^\pm (x)\hat{\a}_j^\pm \rangle =
\pm |\psi_j^\pm (x)\rho_j^\pm \rangle,
\end{gather*}
where
\begin{gather*}
|\psi_j^\pm (x) \rangle = |\psi^\pm (x,\lambda_j^\pm)
\rangle, \qquad |\phi_j^\pm (x) \rangle = |\phi^\pm (x,\lambda_j^\pm)
\rangle, \\
\rho_j^\pm = \hat{\c}_j^\pm \d_j^\pm = \b_j^\pm \hat{\a}_j^\pm,
\qquad \tau_j^\pm = \hat{\a}_j^\pm \b_j^\pm = \d_j^\pm
\hat{\c}_j^\pm
\end{gather*}
and the additional coef\/f\/icients $\b_j^\pm$ and $\d_j^\pm$ are
constant $r\times r$ nondegenerate matrices which, as we shall see
below,  are also part of the minimal sets of scattering data
needed to determine the potential $Q(x,t)$.

These considerations allow us to calculate explicitly the residues
in equations~(\ref{eq:J_11}),  (\ref{eq:J_11t}) with the result
\begin{gather*}
\Res_{\mu =\lambda_j^+} \frac{X^+(x,\mu)}{\mu-\lambda} =
\frac{(|\0\rangle, |\phi_j^+(x)\tau_j^+\rangle)}{\lambda_j^+
-\lambda}, \qquad \Res_{\mu =\lambda_j^+} \frac{\tilde{X}^+(x,\mu)}
{\mu-\lambda} = \frac{(|\psi_j^+(x)\rho_j^+\rangle ,
|\0\rangle)}{\lambda_j^+
-\lambda}, \\
\Res_{\mu =\lambda_j^+} \frac{X^-(x,\mu)}{\mu-\lambda} = -\frac{(
|\phi_j^-(x)\tau_j^-\rangle , |\0\rangle)}{\lambda_j^+ -\lambda},
\qquad \Res_{\mu =\lambda_j^+} \frac{\tilde{X}^-(x,\mu)}
{\mu-\lambda} = -\frac{(|\0\rangle,
|\psi_j^-(x)\tau_j^-\rangle)}{\lambda_j^+ -\lambda},
\end{gather*}
where $|\0\rangle$ stands for a collection of $r$ columns whose
components are all equal to~0.

We can also evaluate $\mathcal{J}_1(\lambda ) $ and
$\mathcal{J}_2(\lambda ) $ by integrating along the contours. In
integrating along the inf\/inite semi-circles of $\gamma
_{\pm,\infty } $ we use the asymptotic behavior of
$X^\pm(x,\lambda ) $ and $\tilde{X}^\pm(x,\lambda ) $ for $\lambda
\to\infty $. The results are
\begin{gather}\label{eq:J_12}
\mathcal{J}_1(\lambda ) = \openone + {1 \over 2\pi \ri}
\int_{-\infty} ^{\infty }{\rd\mu \over \mu -\lambda } \phi(x,\mu )
\re^{\ri \mu Jx}  K(x,\mu), \\
\label{eq:J_12t} \mathcal{J}_2(\lambda ) = \openone + {1 \over
2\pi \ri} \int_{-\infty} ^{\infty }{\rd\mu \over \mu -\lambda }
\psi(x,\mu ) \re^{\ri \mu Jx}  \tilde{K}(x,\mu), \\
K(x,\mu) = \re^{-\ri \mu J x} K_0(\mu)  \re^{\ri \mu J x},
\qquad \tilde{K}(x,\mu) = \re^{-\ri \mu J x} \tilde{K}_0(\mu)  \re^{\ri \mu J x},\nonumber \\
K_0(\mu)= \left(\begin{array}{cc} 0 & \tau^+(\mu)
 \\ \tau^-(\mu)  & 0 \\
\end{array} \right), \qquad \tilde{K}_0(\mu)= \left(\begin{array}{cc}
0 & \rho^+(\mu) \\ \rho^-(\mu)  & 0 \\
\end{array} \right),\nonumber
\end{gather}
where in evaluating the integrands  we made use of equations
(\ref{eq:6.1}), (\ref{eq:6.2}), (\ref{eq:12.4}) and
(\ref{eq:12.5}).

Equating the right hand sides of (\ref{eq:J_11}) and
(\ref{eq:J_12}), and (\ref{eq:J_11t}) and (\ref{eq:J_12t}) we get
the following integral decomposition for $X^\pm(x,\lambda ) $:
\begin{gather}\label{eq:X+}
X^+(x,\lambda ) = \openone + {1 \over 2\pi \ri} \int_{-\infty
}^{\infty } {\rd\mu\over \mu -\lambda } X^-(x,\mu ) K_1(x,\mu ) +
\sum_{j =1}^{N} \frac{X_j^-(x)K_{1,j}(x)}{\lambda _j^- -\lambda},
\\
\label{eq:X-}
X^-(x,\lambda ) = \openone + {1 \over 2\pi \ri} \int_{-\infty
}^{\infty } {\rd\mu\over \mu -\lambda } X^-(x,\mu ) K_2(x,\mu ) -
\sum_{j =1}^{N} \frac{ X_j^+(x)K_{2,j}(x)}{\lambda _j^+ -\lambda},
\end{gather}
where $X^\pm_j(x)=X^\pm(x,\lambda_j^\pm)$ and
\begin{gather*}
K_{1,j}(x) = \re^{-\ri \lambda_j^- J x} \left(\begin{array}{cc} 0
& \rho_j^+ \\ \tau_j^- & 0 \\ \end{array} \right)  \re^{\ri
\lambda_j^- J x},  \qquad K_{2,j}(x) = \re^{-\ri \lambda_j^+ J x}
\left(\begin{array}{cc} 0 & \tau_j^+ \\ \rho_j^- & 0 \\
\end{array} \right)\re^{\ri \lambda_j^+ J x}.
\end{gather*}

Equations (\ref{eq:X+}), (\ref{eq:X-}) can be viewed as a set of
singular integral equations which are equivalent to the RHP. For
the MNLS these were f\/irst derived in \cite{Ma1}.

We end this section by a brief explanation of how the potential
$Q(x,t)$ can be recovered provided we have solved the RHP and know
the solutions $X^\pm(x,\lambda)$. First we take into account that
$X^\pm(x,\lambda)$ satisfy the dif\/ferential equation
\begin{gather}\label{eq:Xpm}
\ri\frac{\rd X^\pm}{\rd x} +Q(x,t)X^\pm(x,\lambda) - \lambda
[J,X^\pm(x,\lambda)] =0
\end{gather}
which must hold true for all $\lambda$. From equation (\ref{eq:12.1})
and also from the integral equations~(\ref{eq:X+}), (\ref{eq:X-})
one concludes that $X^\pm(x,\lambda)$ and their inverse
$\hat{X}^\pm(x,\lambda)$  are regular for $\lambda \to\infty$ and
allow asymptotic expansions of the
form \begin{gather*}
X^\pm(x,\lambda) = \openone
+ \sum_{s=1}^\infty \lambda^{-s}X_s(x), \qquad
\hat{X}^\pm(x,\lambda) = \openone + \sum_{s=1}^\infty \lambda^{-s}
\hat{X}_s(x).
\end{gather*}
Inserting these into equation~(\ref{eq:Xpm}) and taking the limit
$\lambda\to\infty$ we get
\begin{gather}\label{eq:Q'}
Q(x,t) = \lim_{\lambda\to\infty} \lambda (J - X^\pm(x,\lambda) J
\hat{X}^\pm(x,\lambda)]) =[J, X_1(x)].
\end{gather}


\section{The generalized Fourier transforms}\label{sec4}

It is well known that the ISM can be interpreted as a generalized
Fourier~\cite{Varna04} transform which maps the potential $Q(x,t)$
onto the minimal sets of scattering data $\mathcal{T}_i$. Here we
brief\/ly formulate these results and in the next Section we will
analyze how they are modif\/ied under the reduction conditions.

The generalized exponentials are the `squared solutions' which
are determined by the FAS and the Cartan--Weyl basis of the
corresponding algebra as follows
\begin{gather*}
\bPsi_\alpha ^\pm = \chi^\pm(x,\lambda) E_\alpha
\hat{\chi}^\pm(x,\lambda), \qquad \bPhi_\alpha ^\pm =
\chi^\pm(x,\lambda) E_{-\alpha} \hat{\chi}^\pm(x,\lambda), \qquad
\alpha \in \Delta_1^+.
\end{gather*}

\subsection[Expansion over the 'squared solutions']{Expansion over the `squared solutions'}

The `squared solutions' are complete set of functions in the phase
space \cite{Varna04}.  This allows one to expand any function over
the `squared solutions'.

Let us introduce the sets of `squared solutions'
\begin{gather*}
\{\bPsi \} = \{\bPsi \}_{\rm c} \cup \{\bPsi \}_{\rm d}, \qquad
\{\bPhi \} = \{\bPhi \}_{\rm c} \cup \{\bPhi \}_{\rm d},\\
\{\bPsi \}_{\rm c}  \equiv \left\{ \bPsi ^+_{\alpha}(x,\lambda),
\quad \bPsi ^-_{-\alpha}(x,\lambda), \quad i<r, \quad \lambda \in
\bbbr \right\},
\\
\{\bPsi \}_{\rm d}  \equiv \left\{\bPsi ^+_{\alpha;j}(x), \quad
\dot{\bPsi }^+_{\alpha ;j}(x),\quad  \bPsi ^-_{-\alpha;j}(x),\quad
\dot{\bPsi }^-_{-\alpha;j}(x)\right\}_{j=1}^{N},
\\
\{\bPhi \}_{\rm c}  \equiv \left\{ \bPhi ^+_{-\alpha}(x,\lambda),
\quad \bPhi ^-_{\alpha}(x,\lambda), \quad i<r, \quad \lambda \in
\bbbr \right\},
\\
\{\bPhi \}_{\rm d}  \equiv \left\{\bPhi ^+_{-\alpha;j}(x), \quad
\dot{\bPhi }^+_{-\alpha;j}(x),\quad  \bPhi ^-_{\alpha;j}(x),\quad
\dot{\bPhi }^-_{\alpha;j}(x)\right\}_{j=1}^{N},
\end{gather*}
where the subscripts `c' and `d' refer to the continuous and
discrete spectrum of $L $. The `squared solutions' in bold-face
$\bPsi_{\alpha}^{+} $, \dots\ are obtained from $\Psi _{\alpha}^{+}
$, \dots\ by applying the projector $P_{0J} $, i.e.\ $\bPsi
_{\alpha}^{+}(x,\lambda ) = P_{0J}\Psi _{\alpha}^{+}(x,\lambda )
$.

Using the Wronskian relations one can derive the expansions over
the `squared solutions' of two important functions. Skipping the
calculational details  we formulate the results \cite{Varna04}.
The expansion of $Q(x) $ over the systems $ \{\bPhi ^\pm \}$ and $
\{\bPsi ^\pm \}$ takes the form
\begin{gather}
Q(x) = \frac{\ri}{\pi } \int_{-\infty }^{\infty } \rd\lambda
\sum_{\alpha \in \Delta_1^+} \left( \tau^+_{\alpha }(\lambda )
\bPhi_{\alpha } ^+(x, \lambda )
-\tau_{\alpha }^-(\lambda )  \bPhi_{-\alpha } ^-(x, \lambda ) \right) \nonumber\\
\phantom{Q(x) =}{} + 2\sum_{k=1}^{N} \sum_{\alpha \in \Delta_1^+}
\left(\tau^+_{\alpha ;j} \bPhi_{\alpha ;j} ^+(x) + \tau^-_{\alpha
;j} \bPhi_{-\alpha ;j} ^-(x)\right),\label{eq:49.4}
\\
Q(x) =- {\ri \over \pi } \int_{-\infty }^{\infty } \rd\lambda
\sum_{\alpha \in \Delta_1^+} \left( \rho^+_{\alpha }(\lambda )
\bPsi_{-\alpha } ^+(x, \lambda )
-\rho_{\alpha }^-(\lambda )  \bPsi_{\alpha } ^-(x, \lambda ) \right) \nonumber\\
\phantom{Q(x) =}{} - 2\sum_{k=1}^{N} \sum_{\alpha \in \Delta_1^+}
\left(\rho^+_{\alpha ;j} \bPsi_{-\alpha ;j} ^+(x) + \rho^-_{\alpha
;j} \bPsi_{\alpha ;j} ^-(x)\right).\label{eq:49.5}
\end{gather}

The next  expansion is of $\ad_J^{-1}\delta Q(x) $ over the
systems $ \{\bPhi ^\pm \}$ and $ \{\bPsi ^\pm \}$
\begin{gather}
\ad_J^{-1}\delta Q(x) = {\ri\over 2\pi } \int_{-\infty }^{\infty
} \rd\lambda \sum_{\alpha \in \Delta_1^+} \left(
\delta\tau^+_{\alpha }(\lambda ) \bPhi_{\alpha } ^+(x, \lambda )
+ \delta \tau_{\alpha }^-(\lambda )  \bPhi_{-\alpha } ^-(x, \lambda ) \right) \nonumber\\
\phantom{\ad_J^{-1}\delta Q(x) =}{} + \sum_{k=1}^{N} \sum_{\alpha \in \Delta_1^+} \left(\delta
W^+_{\alpha ;j}(x) -
\delta'W^-_{-\alpha ;j}(x) \right), \label{eq:50.6}\\
\ad_J^{-1}\delta Q(x) = {\ri  \over 2\pi } \int_{-\infty
}^{\infty } \rd\lambda \sum_{\alpha \in \Delta_1^+} \left(\delta
\rho^+_{\alpha }(\lambda ) \bPsi_{-\alpha } ^+(x, \lambda )
+ \delta\rho_{\alpha }^-(\lambda )  \bPsi_{\alpha } ^-(x, \lambda ) \right) \nonumber\\
\phantom{\ad_J^{-1}\delta Q(x) =}{} +\sum_{k=1}^{N} \sum_{\alpha \in \Delta_1^+}
\left(\delta\tilde{W}^+_{-\alpha ;j}(x) -
\delta\tilde{W}^-_{\alpha ;j}(x)\right),\label{eq:51.4}
\end{gather}
where
\begin{gather*}
\delta W^\pm_{\pm\alpha ;j}(x)=\delta\lambda_j^\pm
\tau^\pm_{\alpha ;j} \dot{\bPhi}_{\pm\alpha;j} ^\pm(x) + \delta
\tau^\pm_{\alpha ;j} \bPhi_{\pm\alpha;j} ^\pm(x), \\
\delta \tilde{W}^\pm_{\mp \alpha ;j}(x)=\delta\lambda_j^\pm
\rho^\pm_{\alpha;j} \dot{\bPsi}_{\mp\alpha;j} ^\pm (x) + \delta
\rho^\pm_{\alpha;j} \bPsi_{\mp\alpha;j} ^\pm(x)
\end{gather*}
and $\bPhi_{\pm\alpha;j} ^\pm(x)=\bPhi_{\pm\alpha}
^\pm(x,\lambda_j^\pm)$, $\dot{\bPhi}_{\pm\alpha;j} ^\pm(x)
=\partial_\lambda \bPhi_{\pm\alpha} ^\pm(x,\lambda)
|_{\lambda=\lambda_j^\pm} $.

The expansions (\ref{eq:49.4}), (\ref{eq:49.5}) is another way to
establish the one-to-one correspondence bet\-ween~$Q(x) $ and each
of the minimal sets of scattering data $\mathcal{T}_1 $ and
$\mathcal{T}_2 $ (\ref{eq:T_12}). Likewise the expansions
(\ref{eq:50.6}), (\ref{eq:51.4}) establish the one-to-one
correspondence between the variation of the potential $\delta Q(x)
$ and the variations of the scattering data $\delta \mathcal{T}_1
$ and $\delta \mathcal{T}_2 $.

The expansions (\ref{eq:50.6}), (\ref{eq:51.4}) have a special
particular case when one considers the class of variations of
$Q(x,t)$ due to the evolution in $t$. Then
\begin{gather*}
\delta Q(x,t)\equiv Q(x,t+\delta t)-Q(x,t) = \frac{\partial
Q}{\partial t}\delta t + \mathcal(O)((\delta t)^2).
\end{gather*}
Assuming that $\delta t$ is small and keeping only the f\/irst order
terms in $\delta t$ we get the expansions for $\ad_J^{-1}Q_t$.
They are obtained from (\ref{eq:50.6}), (\ref{eq:51.4}) by
replacing $\delta \rho_\alpha^\pm (\lambda)$ and $\delta
\tau_\alpha^\pm (\lambda)$ by $\partial_t \rho_\alpha^\pm
(\lambda)$ and $
\partial_t \rho_\alpha^\pm (\lambda)$.

\subsection{The generating operators}\label{ssec:Lambda}

To complete the analogy between the standard Fourier transform and
the expansions over the `squared solutions' we need the analogs of
the operator $D_0=-\ri \rd/\rd x $. The operator $D_0 $ is the one
for which $\re^{\ri\lambda x} $ is an eigenfunction:  $D_0
\re^{\ri\lambda x}=\lambda \re^{\ri\lambda x} $. Therefore it is
natural to introduce the generating operators $\Lambda _\pm $
through
\begin{gather*}
(\Lambda _+-\lambda )\bPsi_{-\alpha}^{+} (x,\lambda ) = 0, \qquad
(\Lambda _+-\lambda )\bPsi_{\alpha}^{-} (x,\lambda ) = 0, \qquad
(\Lambda _+ -\lambda_j^\pm )\bPsi_{\mp\alpha;j}^{+}(x)= 0, \\
(\Lambda _--\lambda )\bPhi_{\alpha}^{+} (x,\lambda ) = 0, \qquad
(\Lambda _--\lambda )\bPhi_{-\alpha}^{-} (x,\lambda ) = 0, \qquad
(\Lambda _+ -\lambda_j^\pm )\bPhi_{\pm\alpha;j}^{+}(x)= 0,
\end{gather*}
where the generating operators $\Lambda _\pm $ are given by
\begin{gather}\label{eq:**6}
\Lambda _\pm X(x) \equiv \ad_{J}^{-1} \left( \ri {\rd X \over \rd
x} + \ri \left[ Q(x), \int_{\pm\infty }^{x} \rd y\, [Q(y),
X(y)]\right] \right).
\end{gather}

The rest of the squared solutions are not eigenfunctions of
neither $\Lambda _+ $ nor $\Lambda _-$
\begin{gather*}
(\Lambda _+ -\lambda_j^+ )
\dot{\bPsi}_{-\alpha;j}^{+}(x)=\bPsi_{-\alpha;j}^{+}(x), \qquad
(\Lambda _+ -\lambda_j^- ) \dot{\bPsi}_{\alpha;j}^{-}(x)=
\bPsi_{\alpha;j}^{-}(x), \\
(\Lambda _- -\lambda_j^+ )
\dot{\bPhi}_{ir;j}^{+}(x)=\bPhi_{\alpha;j}^{+}(x), \qquad
(\Lambda _- -\lambda_j^- ) \dot{\bPhi}_{\alpha;j}^{-}(x)=
\bPhi_{\alpha;j}^{-}(x),
\end{gather*}
i.e., $\dot{\bPsi}_{\alpha;j}^{+}(x) $ and
$\dot{\bPhi}_{\alpha;j}^{+}(x) $ are adjoint eigenfunctions of
$\Lambda _+ $ and $\Lambda _- $. This means that $\lambda _j^\pm
$, $j=1,\dots, N $ are also the discrete eigenvalues of $\Lambda
_\pm $ but the corresponding eigenspaces of $\Lambda _\pm $ have
double the dimensions of the ones of $L $; now they are spanned
by both $\bPsi_{\mp\alpha;j}^{\pm}(x) $ and
$\dot{\bPsi}_{\mp\alpha;j}^{\pm}(x) $. Thus the sets $\{\Psi \} $
and $\{\Phi \} $ are the complete sets of eigen- and adjoint
functions of $\Lambda _+ $ and $\Lambda _- $.

\subsection{The minimal sets of scattering data}\label{ssec:2x}

Obviously, given the potential $Q(x) $ one can solve the integral
equations for the Jost solutions which determine them uniquely.
The Jost solutions in turn determine uniquely the scattering
matrix $T(\lambda ) $ and its inverse $\hat{T}(\lambda ) $. But
$Q(x) $ contains  $r(r-1) $ independent complex-valued functions
of $x $. Thus it is natural to expect that at most $r(r-1) $ of
the coef\/f\/icients in $T(\lambda ) $ for $\lambda \in \bbbr$ will be
independent; the rest must be functions of those. The set of
independent coef\/f\/icients of $T(\lambda ) $ are known as the
minimal set of scattering data.

The completeness relation for the `squared solutions' ensure that
there is one-to-one correspondence between the potential $Q(x,t)$
and its expansion coef\/f\/icients. Thus we may use as minimal sets
of scattering data the following two sets $\mathcal{T}_{i} \equiv
\mathcal{T}_{i,\rm c}\cup \mathcal{T}_{i,\rm d} $
\begin{gather}
\mathcal{T}_{1,\rm c} \equiv \left\{ \rho ^+(\lambda ), \rho
^-(\lambda ), \quad \lambda \in \bbbr\right\},    \qquad
\mathcal{T}_{1,\rm d} \equiv \left\{\rho_j^\pm, \lambda _j^\pm
\right\}_{j=1}^{N}, \nonumber\\
\mathcal{T}_{2,\rm c} \equiv \left\{ \tau^+(\lambda ),
\tau^-(\lambda ), \quad \lambda \in \bbbr\right\},    \qquad
\mathcal{T}_{1,\rm d} \equiv \left\{\tau_j^\pm, \lambda _j^\pm
\right\}_{j=1}^{N},\label{eq:T_12}
\end{gather}
where the ref\/lection coef\/f\/icients $\rho ^\pm(\lambda ) $ and
$\tau^\pm(\lambda ) $ were introduced in equation (\ref{eq:6.2}),
$\lambda _j^\pm $ are (simple) discrete eigenvalues of $L $ and
$\rho_j^\pm $ and $\tau_j^\pm $ characterize the norming constants
of the corresponding Jost  solutions.

\begin{remark}\label{rem:n2}
A consequence of equation~(\ref{eq:ro-roT}) is the fact that
$\S^\pm(\lambda),\T^\pm(\lambda)\in SO(2r)$. These factors can be
written also in the form
\begin{gather*}
\S^\pm(\lambda) = \exp \left(\sum_{\alpha \in \Delta^+_1}
\tau_\alpha^\pm (\lambda) E_{\pm\alpha} \right), \qquad
\T^\pm(\lambda) = \exp \left(\sum_{\alpha \in \Delta^+_1}
\rho_\alpha^\pm (\lambda) E_{\pm\alpha} \right).
\end{gather*}
Taking into account that in the typical representation we have
$E_{\pm \alpha} E_{\pm \beta}=0$ for all roots $\alpha,\beta \in
\Delta_1^+$ we f\/ind that
\begin{gather}
\sum_{\alpha \in \Delta^+_1} \tau_\alpha^+ (\lambda) E_{\pm\alpha}
= \left( \begin{array}{cc} 0 & \tau^+ (\lambda)
\\ 0 & 0 \\ \end{array} \right) , \qquad  \sum_{\alpha \in \Delta^+_1}
\tau_\alpha^- (\lambda) E_{-\alpha} = \left(\begin{array}{cc} 0 & 0 \\
\tau^-(\lambda) & 0 \\ \end{array}\right), \nonumber \\
\sum_{\alpha \in \Delta^+_1} \rho_\alpha^+(\lambda) E_{\pm\alpha}
= \left( \begin{array}{cc} 0 & \rho^+(\lambda) \\ 0 & 0 \\
\end{array} \right) , \qquad  \sum_{\alpha \in \Delta^+_1}
\rho_\alpha^- (\lambda) E_{-\alpha} =\left(\begin{array}{cc} 0 & 0
\\ \rho^- (\lambda) & 0 \\ \end{array}\right),\label{eq:tau-rho}
\end{gather}
where $\Delta_1^+$ is a subset of the positive roots of $so(2r)$
def\/ined in Subsection~\ref{ssec:2.4}. The formulae~(\ref{eq:tau-rho})
ensure that the number of independent matrix elements of
$\tau^+(\lambda)$ and $\tau^-(\lambda)$ (resp., $\rho^+(\lambda)$
and~$\rho^-(\lambda)$) equals $2|\Delta_1^+| = r(r-1)$ which
coincides with the number of independent functions of~$Q(x)$.
\end{remark}

An important consequence of the expansions is the theorem~\cite{Varna04}

\begin{theorem}\label{th:1}
Any nonlinear evolution equation (NLEE) integrable via the
inverse scattering method applied to the Lax operator $L$
\eqref{eq:LM} can be written in the form
\begin{gather}\label{eq:58.5}
\ri\, {\rm ad}_J^{-1} \frac{\partial Q}{\partial t} + 2 f(\Lambda) Q(x,t) =0,
\end{gather}
where the function $f(\lambda)$ is known as the dispersion law of
this NLEE. The generic MMKdV equation is a member of this class
and is obtained by choosing $f(\lambda)=-4\lambda^3$. If $Q(x,t)$
is a~solution to \eqref{eq:58.5} then the corresponding scattering
matrix satisfy the linear evolution equation
\begin{gather}\label{eq:dTdt}
\ri\frac{dT}{dt} + f(\lambda) [J, T(\lambda,t)]=0,
\end{gather}
or equivalently
\begin{gather*}
\ri {\rd\rho ^\pm \over \rd t} \mp 2f_0(\lambda)\rho ^\pm = 0,
\qquad {\rd\lambda _j^\pm \over \rd t} =0, \qquad \ri
{\rd\rho_{;j} ^\pm \over \rd t} \mp 2f_0(\lambda _j^\pm)\rho_{;j}
^\pm =0,
\\
\ri {d\tau ^\pm \over \rd t} \pm 2f_0(\lambda)\tau ^\pm = 0,
\qquad {\rd\lambda _j^\pm \over \rd t} =0, \qquad i {\rd\tau_{;j}
^\pm \over \rd t} \pm 2f_0(\lambda _j^\pm)\tau_{;j} ^\pm =0,
\end{gather*}
and vice versa. In particular from \eqref{eq:dTdt} there follows
that $\a^\pm(\lambda)$ and $\c^\pm(\lambda)$ are time-independent
and therefore can be considered as generating functionals of
integrals of motion for the NLEE.
\end{theorem}

Let us, before going into the non-trivial reductions, brief\/ly
discuss the Hamiltonian formulations for the generic (i.e.,
non-reduced) MMKdV type equations. It is well known (see
\cite{Varna04} and the numerous references therein) that the class
of these equations is generated by the so-called recursion
operator $\Lambda =1/2(\Lambda_+ + \Lambda_-)$ which is def\/ined by equation~(\ref{eq:**6}).

If no additional reduction is imposed one can write each of the
equations in~(\ref{eq:58.5}) in Hamiltonian form. The
corresponding Hamiltonian and symplectic form for the MMKV equation are
given by
\begin{gather}\label{eq:58.3}
H_{\rm MMKdV}^{(0)} =  \frac{1}{4} \int_{-\infty}^\infty dx\;
\left( \tr(JQ_xQ_{xx}) - 3 \tr (JQ^3Q_x)\right), \\
 \Omega^{(0)} = \frac{1}{\ri} \int_{-\infty}^\infty
dx\; \tr \left( \ad_J^{-1}\delta Q(x) \wedgecomma \left[ J,
\ad_J^{-1}\delta Q(x)\right] \right)  = \frac{1}{2\ri} \int_{-\infty}^\infty dx\; \tr ( J\delta Q(x)
\wedgecomma \delta Q(x)). \nonumber
\end{gather}
The Hamiltonian can be identif\/ied as proportional to the fourth
coef\/f\/icient $I_4$ in the asymptotic expansion of $A^+(\lambda)$
(\ref{eq:2x2-3}) over the negative powers of $\lambda$
\begin{gather*}
A^+(\lambda) = \sum_{k=1}^\infty \ri I_k \lambda^{-k}.
\end{gather*}
This series of integrals of motion is known as the principal one.
The f\/irst three of these integrals take the form
\begin{gather*}
I_1 = {1\over 4 } \int_{-\infty }^{\infty } \rd x\, \tr
(Q^2(x,t)), \qquad I_2 = -{\ri \over 4 } \int_{-\infty }^{\infty }
\rd x\, \tr (Q\ad_J^{-1}Q_x),  \\
I_3 = -{1\over 8 } \int_{-\infty }^{\infty } \rd x\, \tr
(QQ_{xx}+2 Q^4), \qquad I_4=\frac{1}{32}
\int_{-\infty}^\infty dx\; \left( \tr(JQ_xQ_{xx}) - 3 \tr
(JQ^3Q_x)\right).
\end{gather*}

We will remind also another important result, namely that the
gradient of $I_k$ is expressed through $\Lambda$ as
\begin{gather*}
\nabla_{Q^T(x)} I_k = - \frac{1}{2} \Lambda^{k-1} Q(x,t).
\end{gather*}
Then the Hamiltonian equations written through $\Omega^{(0)}$ and
the Hamiltonian vector f\/ield $X_{H^{(0)}}$ in the form
\begin{gather}\label{eq:58.7}
\Omega^{(0)} (\cdot , X_{H^{(0)}} ) + \delta H^{(0)} =0
\end{gather}
for $H^{(0)}$ given by (\ref{eq:58.3}) coincides with the MMKdV
equation.

An alternative way to formulate Hamiltonian equations of motion is
to introduce along with the Hamiltonian the Poisson brackets on
the phase space $\mathcal{M}$ which is the space of smooth
functions taking values in $\mathfrak{g}^{(0)}$ and vanishing fast
enough for $x\to\pm\infty$, see (\ref{eq:Q}). These  brackets can
be introduced by
\begin{gather*}
\{ F, G\}_{(0)} = \ri \int_{-\infty}^\infty dx\; \tr \left(
\nabla_{Q^T(x)} F, \left[ J, \nabla_{Q^T(x)} G\right] \right) .
\end{gather*}
Then the Hamiltonian equations of motions
\begin{gather}\label{eq:59.0}
\frac{dq_{ij}}{dt} = \{ q_{ij}, H^{(0)}\}_{(0)}, \qquad
\frac{dp_{ij}}{dt} = \{ p_{ij}, H^{(0)}\}_{(0)}
\end{gather}
with the above choice for $ H^{(0)} $ again give the MMKdV
equation.

Along with this standard Hamiltonian formulation there exist a
whole hierarchy of them. This is a special property of the
integrable NLEE. The hierarchy is generated again by the recursion
operator and has the form
\begin{gather*}
H_{\rm MMKdV}^{(m)} = - 8 I_{4+m}, \qquad \Omega^{(m)} =
\frac{1}{\ri} \int_{-\infty}^\infty dx\; \tr \left( \ad_J^{-1}\delta
Q(x) \wedgecomma \left[ J, \Lambda^m \ad_J^{-1}\delta Q(x)\right]
\right) .
\end{gather*}
Of course there is also a hierarchy of Poisson brackets
\begin{gather*}
\{ F, G\}_{(m)} = \ri \int_{-\infty}^\infty dx\; \tr \left(
\nabla_{Q^T(x)} F, \left[ J, \Lambda^{-m} \nabla_{Q^T(x)} G\right]
\right).
\end{gather*}
For a f\/ixed value of $m$ the Poisson bracket $\{ \cdot , \cdot
\}_{(m)}$ is dual to the symplectic form $\Omega^{(m)}$ in the
sense that combined with a given Hamiltonian they produce the same
equations of motion. Note that since $\Lambda$ is an
integro-dif\/ferential operator in general it is not easy to
evaluate explicitly its negative powers. Using this duality one
can avoid the necessity to evaluate negative powers of~$\Lambda$.

Then the analogs of (\ref{eq:58.7}) and (\ref{eq:59.0}) take the
form:
\begin{gather}\label{eq:58.7m}
\Omega^{(m)} (\cdot , X_{H^{(m)}} ) + \delta H^{(m)} =0,
\\
\label{eq:59.0m}
\frac{dq_{ij}}{dt} = \{ q_{ij}, H^{(-m)}\}_{(m)}, \qquad
\frac{dp_{ij}}{dt} = \{ p_{ij}, H^{(-m)}\}_{(m)},
\end{gather}
where the hierarchy of Hamiltonians is given by:
\begin{gather}\label{eq:59.3}
H^{(m)} = -4 \sum_k f_kI_{k+1-m}.
\end{gather}
The equations (\ref{eq:58.7m}) and (\ref{eq:59.0m}) with the
Hamiltonian $H^{(m)}$ given by (\ref{eq:59.3}) will produce the
NLEE (\ref{eq:58.5}) with dispersion law $f(\lambda) = \sum_k
f_k\lambda^k$ for any value of $m$.

\begin{remark}\label{rem:n3}
It is a separate issue to prove that the hierarchies of symplectic
structures and Poisson brackets have all the necessary properties.
This is done using the spectral decompositions of the recursion
operators $\Lambda_\pm$ which are known also as the expansions
over the `squared solutions' of $L$. We refer the reader to the
review papers \cite{G-cm,Varna04} where he/she can f\/ind the proof
of the completeness relation for the `squared solutions' along
with the proof that any two of the symplectic forms introduced
above are compatible.

\end{remark}

\subsection{The reduction group of Mikhailov}\label{sec4.4}

The reduction group $G_R $ is a f\/inite group which preserves the
Lax representation (\ref{eq:LM}), i.e.\ it ensures that the
reduction constraints are automatically compatible with the
evolution. $G_R $~must have two realizations: i) $G_R \subset {\rm
Aut}\,\mathfrak{g} $ and ii) $G_R \subset {\rm Conf}\, \Bbb C $,
i.e.\ as conformal mappings of the complex $\lambda $-plane. To
each $g_k\in G_R $ we relate a reduction condition for the Lax
pair as follows~\cite{Mikh}
\begin{gather}\label{eq:2.1}
C_k(L(\Gamma _k(\lambda ))) = \eta _k L(\lambda ), \qquad
C_k(M(\Gamma _k(\lambda ))) = \eta _k M(\lambda ),
\end{gather}
where $C_k\in \mbox{Aut}\; \mathfrak{g} $ and $\Gamma _k(\lambda
)\in \mbox{Conf\,} \bbbc $ are the images of $g_k $ and $\eta _k
=1 $ or $-1 $ depending on the choice of $C_k $. Since $G_R $ is a
f\/inite group then for each $g_k $ there exist an integer $N_k $
such that $g_k^{N_k} =\openone $.

More specif\/ically the automorphisms $C_k $, $k=1,\dots,4 $ listed
above lead to the following reductions for the potentials
$U(x,t,\lambda )$ and $V(x,t,\lambda )$ of the Lax pair
\begin{gather*}
U(x,t,\lambda ) = Q(x,t) - \lambda J, \qquad V(x,t,\lambda ) =
\sum_{k=0}^2  \lambda^k V_k(x,t) -4\lambda^3J,
\end{gather*}
of the Lax representation
\begin{alignat}{4}\label{eq:U-V.a}
& \mbox{1)} \quad && C_1(U^{\dagger}(\kappa _1(\lambda )))=
U(\lambda ),
\qquad & & C_1(V^{\dagger}(\kappa _1(\lambda )))= V(\lambda ),&  \\
\label{eq:U-V.b}  & \mbox{2)} \quad &&  C_2(U^{T}(\kappa _2(\lambda
)))= -U(\lambda ), \qquad
&& C_2(V^{T}(\kappa _2(\lambda )))= -V(\lambda ), & \\
\label{eq:U-V.c} & \mbox{3)} \quad && C_3(U^{*}(\kappa _1(\lambda
)))= -U(\lambda ), \qquad
&& C_3(V^{*}(\kappa _1(\lambda )))= -V(\lambda ), & \\
\label{eq:U-V.d} &  \mbox{4)}\quad &&  C_4(U(\kappa _2(\lambda )))=
U(\lambda ), \qquad && C_4(V(\kappa _2(\lambda )))= V(\lambda ),&
\end{alignat}
where
\begin{gather*}
\mbox{a)}\quad \kappa_1(\lambda) =\lambda^*, \qquad \mbox{b)}\quad
\kappa_2(\lambda) =-\lambda.
\end{gather*}
The condition (\ref{eq:2.1}) is obviously compatible with the
group action.

\section{Finite order reductions of MMKdV equations}\label{sec5}

In order that the potential $Q(x,t)$ be relevant for a {\bf DIII}
type symmetric space it must be of the form
\begin{gather*}
Q(x,t) = \sum_{\alpha\in\Delta_1^+} \left( q_\alpha (x,t)
E_{\alpha}+  p_\alpha (x,t) E_{-\alpha} \right),
\end{gather*}
or, equivalently
\begin{gather}\label{eq:Q}
Q(x,t) = \sum_{1\leq i<j \leq r} \left( q_{ij}(x,t) E_{e_i+e_j} +
p_{ij}(x,t) E_{-e_i-e_j} \right) .
\end{gather}

In the next two subsections we  display new reductions of the
MMKdV equations.

\subsection[Class A Reductions preserving $J$]{Class A Reductions preserving $\boldsymbol{J}$}

The class A reductions can be applied also to the MMKdV type
equations. The corresponding automorphisms $C$ preserve $J$, i.e.\
$C^{-1}JC =J$ and are of the form
\begin{gather*}
    C^{-1} U^\dag (x,\lambda^*) C = U(x,\lambda), \qquad
     U(x,\lambda)= Q(x,t) -\lambda J,
\end{gather*}
where $J$ is an element of the Cartan subalgebra dual to the
vector $e_1+e_2+e_3+e_4$. In the typical representation of $so(8)$
$U(x,\lambda)$ takes the form
\begin{gather*}
U(x,t,\lambda) = \left(\begin{array}{cc}
\lambda \openone & q(x,t) \\  p(x,t) & -\lambda \openone \\
\end{array}\right), \qquad q(x,t) = \left(\begin{array}{cccc}
q_{14} & q_{13} & q_{12} & 0 \\ q_{24} & q_{23} & 0 & q_{12} \\
q_{34} & 0 & q_{23} & -q_{13} \\ 0 & q_{34} & -q_{24} & q_{14} \\
\end{array}\right), \nonumber\\ p(x,t) = \left(\begin{array}{cccc}
p_{14} & p_{24} & p_{34} & 0 \\ p_{13} & p_{23} & 0 & p_{34} \\
p_{12} & 0 & p_{23} & -p_{24} \\ 0 & p_{12} & -p_{13} & p_{14} \\
\end{array}\right).
\end{gather*}

\begin{remark}\label{rem:J1}
The automorphisms that satisfy $C^{-1}JC =J$ naturally preserve
the eigensubspaces of $\ad_J$; in other words their action on the
root space maps the subsets of roots $\Delta_1^\pm$ onto
themselves:  $C\Delta_1^\pm =\Delta_1^\pm$.
\end{remark}

We list here several inequivalent reductions of the  Zakharov--Shabat system. In
the f\/irst one we choose $C=C_0$ to be an element of the Cartan
subgroup
\begin{gather*}
    C_0 = \exp \left(\pi \ri \sum_{k=1}^4 s_k h_k \right),
\end{gather*}
where $s_k$ take the values $0$ and $1$. This condition means that
$C_0^2=\openone$, so this will be a $\bbbz_2$-reduction, or
involution. Then the f\/irst example of $\bbbz_2$-reduction  is
\begin{gather*}
C_0^{-1} Q^\dag (x,t) C_0 =Q(x,t),
\end{gather*}
or in components
\begin{gather*}
p_{ij}=\epsilon_{ij} q_{ij}^*, \qquad \epsilon_{ij} =
\epsilon_i\epsilon_j, \qquad \epsilon_j = e^{\pi i s_j}=\pm 1.
\end{gather*}
Obviously $\epsilon_j$ takes values $\pm 1$ depending on whether
$s_j$ equals $0$ or $1$.

The next examples of $\bbbz_2$-reduction correspond to several
choices of $C$ as elements of the Weyl group eventually combined
with the Cartan subgroup element $C_0$
\begin{gather*}
    C_1 = S_{e_1-e_2} S_{e_3-e_4} C_0,
\end{gather*}
where $S_{e_i-e_j}$ is the Weyl ref\/lection related to the root
$e_i-e_j$. Again we have a $\bbbz_2$-reduction, or an involution
\begin{gather*}
C_1^{-1} Q^\dag (x,t) C_1 =Q(x,t).
\end{gather*}
Written in components it takes the form ($\epsilon_{12}=
\epsilon_{34}=1$)
\begin{gather*}
p_{12} = - q_{12}^*, \qquad p_{24} =-\epsilon_{23} q_{13}^*,
\qquad p_{23}=-\epsilon_{13} q_{14}^*,\\
p_{14} = -\epsilon_{13} q_{23}^*, \qquad p_{13} =-\epsilon_{23}
q_{24}^*, \qquad p_{34}=- q_{34}^*.
\end{gather*}
The corresponding Hamiltonian and symplectic form take the form
\begin{gather*}
 8 I_3 
 =-\int_{-\infty}^{\infty} \Big(\partial_{x}q_{12}^{*} \partial_{x}q_{12}+
\partial_{x}q_{34}^{*}\partial_{x}q_{34} +
\epsilon_{13}(q_{23}^{*}\partial_{x}q_{14}+\partial_{x}q_{14}^{*}
\partial_{x}q_{23})   \\
\phantom{8 I_3=}{}
+ \epsilon_{23}\big(\partial_{x}q_{24}^{*}\partial_{x}q_{13}
+\partial_{x}q_{13}^{*}\partial_{x}q_{24}) \Big) d x
+\int_{-\infty}^{\infty} \Big(\epsilon_{12}q_{12}^{*}q_{12}
+q_{34}^{*}q_{34}   \\
\phantom{8 I_3=}{} + \epsilon_{13}(q_{23}^{*}q_{14} +q_{24}^{*}q_{13}
) + \epsilon_{23}(q_{14}^{*}q_{23}+ q_{13}^{*}q_{24})\Big)^2 d x \\
\phantom{8 I_3=}{} + \int_{-\infty}^{\infty}
\left|q_{13}q_{24}+q_{12}q_{34}-q_{14}q_{23}\right|^2 d x,
\\
\Omega^{(0)} = \frac{1}{\ri} \int_{-\infty}^{\infty} \Big(\delta
q_{12}^{*}\wedge \delta q_{12}+ \epsilon_{13}(q_{23}^{*}\wedge
\delta q_{14}+\delta q_{14}^{*}
\wedge\delta q_{23}) \\
 \phantom{\Omega^{(0)} =}{} + \epsilon_{23}(\delta q_{24}^{*}\wedge \delta q_{13}
+\delta q_{13}^{*} \wedge \delta q_{24}) + \delta q_{34}^{*}\wedge
\delta q_{34}  \Big) d x .
\end{gather*}

Another inequivalent examples of $\bbbz_2$-reduction corresponds
to
\begin{gather*}
    C_2 = S_{e_1-e_2} C_0.
\end{gather*}
The involution is $C_2^{-1} Q^\dag (x,t) C_2 =Q(x,t)$, or in
components it takes the form
\begin{gather*}
p_{12} = -\epsilon_{12} q_{12}^*, \qquad p_{24}=-\epsilon_{13}
q_{14}^*, \qquad p_{23}=-\epsilon_{14} q_{13}^*,\\
p_{14} = -\epsilon_{23} q_{24}^*, \qquad p_{13}=-\epsilon_{24}
q_{23}^*, \qquad p_{34}=-\epsilon_{34} q_{34}^*.
\end{gather*}
As a consequence we get
\begin{gather*}
  8 I_3 =-\int_{-\infty}^{\infty} \Big(\epsilon_{12}
\partial_{x}q_{12}^{*} \partial_{x}q_{12}+
\epsilon_{34}\partial_{x}q_{34}^{*}\partial_{x}q_{34} +
\epsilon_{14}\partial_{x}q_{13}^{*} \partial_{x}q_{23}
\\  \phantom{8 I_3 =}{}  +
\epsilon_{23} \partial_{x} q_{24}^{*}\partial_{x}q_{14} +
\epsilon_{24}\partial_{x}q_{23}^{*} \partial_{x}q_{13} +
\epsilon_{13}\partial_{x}q_{14}^{*}\partial_{x}q_{24} \Big) d
x   \\
\phantom{8 I_3 =}{} +\int_{-\infty}^{\infty} \Big(\epsilon_{12}|q_{12}|^2 +
\epsilon_{34} |q_{34}|^2 + \epsilon_{23}q_{24}^{*}q_{14} +
\epsilon_{24} q_{23}^{*}q_{13} + \epsilon_{14} q_{13}^{*}q_{23}+ \epsilon_{13}q_{14}^{*}q_{24}\Big)^2 d x   \\
\phantom{8 I_3 =}{} +
\epsilon_{12}\epsilon_{34} \int_{-\infty}^{\infty}
\left|q_{13}q_{24}+q_{12}q_{34}-q_{14}q_{23}\right|^2 d x,
\\
\Omega^{(0)} = \frac{1}{\ri} \int_{-\infty}^{\infty}
\Big(\epsilon_{12} \delta q_{12}^{*}\wedge \delta q_{12}+
\epsilon_{34}q_{34}^{*}\wedge \delta q_{34}+ \epsilon_{23}\delta
q_{24}^{*} \wedge\delta q_{14}  \\
 \phantom{\Omega^{(0)} =}{}  + \epsilon_{14}\delta q_{13}^{*}\wedge \delta q_{23}
+\epsilon_{24}\delta q_{23}^{*} \wedge \delta q_{13} +
\epsilon_{13}\delta q_{14}^{*}\wedge \delta q_{24} \Big) d x .
\end{gather*}

Next we consider a $\bbbz_3$-reduction generated by $C_3=
S_{e_1-e_2} S_{e_2-e_3} $ which also maps $J$ into $J$. It splits
each of the sets $\Delta_1^\pm$ into two orbits which  are
\begin{gather*}
\mathcal(O)_1^\pm =\{ \pm(e_1+e_2) \;  \pm (e_2+e_3) \; \pm
(e_1+e_3)\}, \\ \mathcal(O)_2^\pm =\{ \pm (e_1+e_4) \; \pm
(e_2+e_4) \; \pm (e_3+e_4) \}.
\end{gather*}

In order to be more ef\/f\/icient we make use of the following basis
in $\mathfrak{g}^{(0)}$
\begin{gather*}
\mathcal{E}_\alpha^{(k)} = \sum_{p=0}^2 \omega^{-kp} C_3^{-k}
E_\alpha C_3^k, \qquad \mathcal{F}_\alpha^{(k)} = \sum_{p=0}^2
\omega^{-kp} C_3^{-k} F_\alpha C_3^k,
\end{gather*}
where $\omega = \exp(2\pi i/3)$ and $\alpha $ takes values
$e_1+e_2$ and $e_1+e_4$. Obviously
\begin{gather}\label{eq:C3}
C_3^{-1} \mathcal{E}_\alpha^{(k)} C_3 =\omega^k
\mathcal{E}_\alpha^{(k)}, \qquad C_3^{-1} \mathcal{F}_\alpha^{(k)}
C_3 =\omega^k \mathcal{F}_\alpha^{(k)}.
\end{gather}
In addition, since $\omega^*=\omega^{-1}$ we get
$(\mathcal{E}_\alpha^{(0)})^\dag = \mathcal{F}_\alpha^{(0)}$ and
$(\mathcal{E}_\alpha^{(k)})^\dag = \mathcal{F}_\alpha^{(3-k)}$ for
$k=1,2$. Then we introduce the potential
\begin{gather*}
Q(x,t)= \sum_{k=0}^3 \sum_{\alpha } \left( q_\alpha^{(k)}(x,t)
\mathcal{E}_\alpha^{(k)} + p_\alpha^{(k)}(x,t)
\mathcal{F}_\alpha^{(k)} \right).
\end{gather*}
In view of equation (\ref{eq:C3}) the reduction condition
(\ref{eq:U-V.a}) leads to the following relations between the
coef\/f\/icients
\begin{gather}
 p_{12}^{(0)} = (q^{(0)}_{12})^*, \qquad p_{12}^{(k)} = \omega^k
(q^{(3-k)}_{12})^*, \qquad q_{12}^{(k)} = \omega^k
(p^{(3-k)}_{12})^*,\nonumber \\
 p_{14}^{(0)} = (q^{(0)}_{14})^*, \qquad p_{14}^{(k)} =\omega^k
(q^{(3-k)}_{14})^*, \qquad q_{14}^{(k)} = \omega^k
(p^{(3-k)}_{14})^*,\label{eq:conZ3}
\end{gather}
where $k=1,2$. It is easy to check that from the conditions
(\ref{eq:conZ3}) there follows $p_{12}^{(k)} =
q_{12}^{(k)}=p_{14}^{(k)} = q_{14}^{(k)}=0$. So we are left with
only one pair of independent functions $q_{12}^{(0)}$ and $
q_{14}^{(0)}$ and their complex conjugate $p_{14}^{(0)}$,
$q_{14}^{(0)}$.

Similarly the reduction (\ref{eq:U-V.b}) leads to
\begin{gather}
 q_{12}^{(0)} = -(q^{(0)}_{12})^*, \qquad q_{12}^{(k)} =
-\omega^{3-k} (q^{(3-k)}_{12})^*, \qquad q_{14}^{(k)} =
-\omega^{3-k} (q^{(3-k)}_{14})^*,\nonumber \\
 p_{12}^{(0)} = -(p^{(0)}_{12})^*, \qquad p_{12}^{(k)} =
-\omega^{3-k} (p^{(3-k)}_{12})^*, \qquad p_{14}^{(k)} =
-\omega^{3-k} (p^{(3-k)}_{14})^* , \label{eq:conZ3b}
\end{gather}
where $k=1,2$. Again from the conditions (\ref{eq:conZ3b}) there
follows $p_{12}^{(k)} = q_{12}^{(k)}=p_{14}^{(k)} =
q_{14}^{(k)}=0$. So we are left with  two pairs of purely
imaginary independent functions: $q_{12}^{(0)}$, $ q_{14}^{(0)}$
and  $p_{12}^{(0)}$, $q_{14}^{(0)}$.

The corresponding Hamiltonian and symplectic form are obtained
from the slightly more general formulae below by imposing the
constraints (\ref{eq:conZ3}) and (\ref{eq:conZ3b}). Here for
simplicity we skip the upper zeroes in $q_{ij}$ and $p_{ij}$
\begin{gather*}
H_{\rm MMKdV} = \frac{1}{6}\int_{-\infty}^{\infty} d x\,
\left(\partial^2_{x}q_{12} \partial_{x}p_{12} -
\partial_{x}q_{12} \partial^2_{x}p_{12} + \partial^2_{x}q_{14}
\partial_{x}p_{14} - \partial_{x}q_{14} \partial^2_{x}p_{14}\right)
\nonumber \\
\phantom{H_{\rm MMKdV} =}{} - \frac{1}{12}\int_{-\infty}^{\infty} \left(p_{12}^2
q_{12}^{2}\partial_{x} -p_{12}^2
\partial_{x}q_{12}^2 + q_{14}^{2}\partial_{x}p_{14}^2
- p_{12}^2 \partial_{x}q_{12}^2 \right) d x,
\\
8I_3=\frac{4}{3}\int_{-\infty}^{\infty} d x\, \left(
\partial_{x}q_{12}\partial_{x}p_{12} +
\partial_{x}q_{14}\partial_{x}p_{14}\right) -\frac{8}{9}
\int_{-\infty}^{\infty} \left(q_{14}^{2}p_{14}^2
+q_{12}^{2}p_{12}^2 \right) d x,
\\
\Omega^{(0)} =\frac{4}{3}\int_{-\infty}^{\infty} d x\,
\left(\delta q_{14}\wedge \delta p_{14} +\delta q_{12}\wedge
\delta p_{12}\right),
\end{gather*}
i.e.\ in this case we get two decoupled mKdV equations.

The $\bbbz_4$-reduction generated by $C_4=S_{e_1-e_2} S_{e_2-e_3}
S_{e_3-e_4}$  also maps $J$ into $J$. It splits each of the sets
$\Delta_1^\pm$ into two orbits which  are
\begin{gather*}
\mathcal(O)_1^\pm =\{ \pm(e_1+e_2)    \pm (e_2+e_3)   \pm
(e_3+e_4)    \pm (e_1+e_4) \}, \\ \mathcal(O)_2^\pm  =\{ \pm
(e_1+e_3)   \pm (e_2+e_4) \}.
\end{gather*}
Again we make use of a convenient basis in $\mathfrak{g}^{(0)}$
\begin{gather*}
\mathcal{E}_\alpha^{(k)} = \sum_{p=0}^3 i^{-kp} C_4^{-k} E_\alpha
C_4^k, \qquad \mathcal{F}_\alpha^{(k)} = \sum_{p=0}^3 i^{-kp}
C_4^{-k} F_\alpha C_4^k,
\end{gather*}
where $\alpha $ takes values $e_1+e_2$ and $e_1+e_3$. Obviously
\begin{gather}\label{eq:C4}
C_4^{-1} \mathcal{E}_\alpha^{(k)} C_4 =i^k
\mathcal{E}_\alpha^{(k)}, \qquad C_4^{-1} \mathcal{F}_\alpha^{(k)}
C_4 =i^k \mathcal{F}_\alpha^{(k)}
\end{gather}
and in addition, $(\mathcal{E}_\alpha^{(0)})^\dag =
\mathcal{F}_\alpha^{(0)}$, $(\mathcal{E}_\alpha^{(k)})^\dag =
\mathcal{F}_\alpha^{(4-k)}$ and $(\mathcal{E}_\alpha^{(k)})^* =
\mathcal{E}_\alpha^{(4-k)}$ for $k=1,2,3$. Then we introduce the
potential
\begin{gather*}
Q(x,t)= \sum_{k=0}^3 \sum_{\alpha } \left( q_\alpha^{(k)}(x,t)
\mathcal{E}_\alpha^{(k)} + p_\alpha^{(k)}(x,t)
\mathcal{F}_\alpha^{(k)} \right).
\end{gather*}
In view of equation (\ref{eq:C4}) the reduction condition
(\ref{eq:U-V.a}) leads to the following relations between the
coef\/f\/icients
\begin{gather}\label{eq:conZ4}
p_{\alpha}^{(0)} = (q^{(0)}_{\alpha})^*, \qquad p_{\alpha}^{(k)} =
i^k(q^{(4-k)}_{\alpha})^*, \qquad q_{\alpha}^{(k)} = i^k
(p^{(4-k)}_{\alpha})^* 
\end{gather}
for $k=1,2,3$. Here $p_\alpha$, $q_\alpha$ coincide with $p_{12}$,
$q_{12}$ (resp.~$p_{13}$, $q_{13}$) for $\alpha=e_1+e_2$ (resp.~$\alpha=e_1+e_3$). Analogously the reduction  (\ref{eq:U-V.b})
gives
\begin{gather}
q_{\alpha}^{(0)} = -(q^{(0)}_{\alpha})^*, \qquad
p_{\alpha}^{(0)} = -(p^{(0)}_{\alpha})^*, \qquad  q_{\alpha}^{(k)} =
-i^k(q^{(4-k)}_{\alpha})^*, \qquad  p_{\alpha}^{(k)} = -i^k
(p^{(4-k)}_{\alpha})^*  \!\!\!\label{eq:conZ4'}
\end{gather}
for $k=1,2,3$. Both conditions (\ref{eq:conZ4}), (\ref{eq:conZ4'})
lead to $p_{12}^{(k)} = q_{12}^{(k)}=p_{14}^{(k)} =
q_{14}^{(k)}=0$ for $k=1,3$. In addition it comes up that
$\mathcal{E}_{13}^{(0)} = \mathcal{E}_{13}^{(2)}
=\mathcal{F}_{13}^{(0)}= \mathcal{F}_{13}^{(2)}$. So we are left
with only two pairs of independent functions $p_{12}^{(0)}$, $
q_{12}^{(0)}$ and $p_{12}^{(2)}$, $ q_{12}^{(2)}$. We provide
below slightly more general formulae for the corresponding
Hamiltonian and symplectic form which are obtained by imposing the
constraints (\ref{eq:conZ4}) or (\ref{eq:conZ4'});  again for
simplicity of notations we skip the upper zeroes in $q_{ij}^{(0)}$
and~$p_{ij}^{(0)}$ and replace $q^{(2)}_{ij}$ and~$p^{(2)}_{ij}$
by $\tilde{q}_{ij}$ and $\tilde{p}_{ij}$
\begin{gather*}
H_{\rm MMKdV} =  \frac{1}{4}\int_{-\infty}^{\infty} d x\,
\left(\partial^2_{x}q_{12} \partial_{x}p_{12} -
\partial_{x}q_{12} \partial^2_{x}p_{12} +
\partial^2_{x}\tilde{q}_{12}
\partial_{x}\tilde{p}_{12} - \partial_{x}\tilde{q}_{12}
\partial^2_{x}\tilde{p}_{12}\right)  \\
\phantom{H_{\rm MMKdV} =}{}  - \frac{3}{32}\int_{-\infty}^{\infty} \Big( \left(
\partial_x (p_{12}^2) + \partial_x (\tilde{p}_{12}^2)\right)
\left( q_{12}^{2} + \tilde{q}_{12}^{2} \right)
 + p_{12} \tilde{p}_{12} \partial_x \left(q_{12}\tilde{q}_{12}\right)
   \\
\phantom{H_{\rm MMKdV} =}{}   + \left(\partial_x (q_{12}^2) + \partial_x
(\tilde{q}_{12}^2)\right) \left( p_{12}^{2} + \tilde{p}_{12}^{2}
\right) + q_{12} \tilde{q}_{12} \partial_x
\left(p_{12}\tilde{p}_{12}\right) \Big) d x,
\\
8I_3 =2\int_{-\infty}^{\infty} \left( \partial_{x}q_{12}
\partial_{x}p_{12} + \partial_{x}\tilde{q}_{12}
\partial_{x}\tilde{p}_{12}\right)  d x \\
\phantom{8I_3 =}{}  -\int_{-\infty}^{\infty} \left(  \left(q_{12}p_{12}
+\tilde{q}_{12} \tilde{p}_{12} \right)^2 + (q_{12}\tilde{p}_{12}
+\tilde{q}_{12} p_{12})^2 \right) dx,
\\
\Omega^{(0)} =2\int_{-\infty}^{\infty} d x\, \left(\delta
q_{12}\wedge \delta p_{12} +\delta \tilde{q}_{12}\wedge \delta
\tilde{p}_{12}\right).
\end{gather*}
Now we get two specially coupled mKdV-type equations. Now we get
four specially coupled mKdV-type equations given by
\begin{gather*}
\partial_{t} q_{0} +\partial_{x}^{3} q_{0}+ \frac{3}{2}( q_{2}
p_{2} +  q_{0} p_{0})\partial_x q_{0} + \frac{3}{2} (q_{2} p_{0} +
q_{0} p_{2})\partial_x q_{2} =0,\\
\partial_{t} q_{2} +\partial_{x}^{3} q_{2}+ \frac{3}{2} (q_{2}
p_{0}+ q_{0} p_{2})\partial_x q_{0} + \frac{3}{2} (q_{2} p_{2}+
q_{0} p_{0})\partial_x q_{2} =0,\\
\partial_{t} p_{0} +\partial_{x}^{3} p_{0}+ \frac{3}{2} (q_{2}
p_{2}+ q_{0} p_{0})\partial_x p_{0} +\frac{3}{2} (q_{0} q_{2}+
q_{2} p_{0})\partial_x p_{2} =0,\\
\partial_{t} p_{2} +\partial_{x}^{3} p_{2}+ \frac{3}{2} (q_{2}
p_{0}+  q_{0} p_{2})\partial_x p_{0} + \frac{3}{2} (q_{0} p_{0}+
q_{2} p_{2})\partial_x p_{2} =0,
\end{gather*}
where we use for simplicity
\begin{gather*}
q_{12}=q_{0},\qquad \tilde{q}_{12}=q_{2},\qquad
p_{12}=p_{0},\qquad \tilde{p}_{12}=p_{2}
\end{gather*}
and with second reduction (\ref{eq:conZ4'}) $p_{0}=q_{0}^*$ and
$p_{2}=-q_{2}^{*}$ we have
\begin{gather}
\partial_{t} q_{0} +\partial_{x}^{3} q_{0}+ \frac{3}{2}(-q_{2}
q^{*}_{2} +  q_{0} q_{0}^{*})\partial_x q_{0} + \frac{3}{2} (q_{2}
q_{0}^{*} - q_{0} q^{*}_{2})\partial_x q_{2} =0,\nonumber\\
\partial_{t} q_{2} +\partial_{x}^{3} q_{2}+ \frac{3}{2} (q_{2}
q^{*}_{0}- q_{0} q_{2}^{*})\partial_x q_{0} + \frac{3}{2} (q_{2}
q_{2}^{*}- q_{0} q_{2}^{*})\partial_x q_{2} =0. \label{C61-C6Z2}
\end{gather}
Obviously in the system (\ref{C61-C6Z2}) we can put both
$q_{0}$, $q_{2}$ real with the result
\begin{gather*}
\partial_{t} q_{0} +\partial_{x}^{3} q_{0}+ \frac{3}{2}(
q_{0}^{2}-q_{2}^2)\partial_x q_{0}=0, \qquad
\partial_{t} q_{2} +\partial_{x}^{3} q_{2} + \frac{3}{2}
(q_{0}^{2}-q_{2}^{2})\partial_x q_{2}=0.
\end{gather*}

The  reduction (\ref{eq:U-V.b}) means that
$q_\alpha^{(0)}=iq^{\vee}_{0}$, $p_\alpha^{(0)}=ip^{\vee}_{0}$,
$q_\alpha^{(2)}=q^{\vee}_{2}$, $p_\alpha^{(2)}=p^{\vee}_{2}$ with
real valued $p^{\vee}_{i}$, $q^{\vee}_{i}$, $i=0,2$. Thus we get
\begin{gather*}
\partial_{t} q^{\vee}_{0} +\partial_{x}^{3} q^{\vee}_{0}+
\frac{3}{2} (q^{\vee}_{2} p^{\vee}_{0}+ q^{\vee}_{0} p^{\vee}_{2})
\partial_x q^{\vee}_{2} + \frac{3}{2} (p^{\vee}_{2}
q^{\vee}_{2}- q^{\vee}_{0} p^{\vee}_{0})\partial_x q^{\vee}_{0}
=0,\\
\partial_{t} q^{\vee}_{2} +\partial_{x}^{3} q^{\vee}_{2}+ \frac{3}{2} (q^{\vee}_{2}
p^{\vee}_{2}- q^{\vee}_{0} p^{\vee}_{0})\partial_x q^{\vee}_{2} +
\frac{3}{2} (q^{\vee}_{2} p^{\vee}_{0}+ q^{\vee}_{0}
p^{\vee}_{2})\partial_x q^{\vee}_{0} =0, \\
\partial_{t} p^{\vee}_{0} +\partial_{x}^{3} p^{\vee}_{0}+ \frac{3}{2} (q^{\vee}_{2}
p^{\vee}_{0}-q^{\vee}_{0} p^{\vee}_{2})\partial_x p^{\vee}_{0} +
\frac{3}{2} (-q^{\vee}_{0} p^{\vee}_{0}+ q^{\vee}_{2}
p^{\vee}_{2})\partial_x p^{\vee}_{2} =0, \\
\partial_{t} p^{\vee}_{2} +\partial_{x}^{3} p^{\vee}_{2}+ \frac{3}{2} (q^{\vee}_{2}
p^{\vee}_{2}- q^{\vee}_{0} p^{\vee}_{0})\partial_x p^{\vee}_{0} +
\frac{3}{2} (q^{\vee}_{0} p^{\vee}_{2}- q^{\vee}_{2}
p^{\vee}_{0})\partial_x p^{\vee}_{2} =0.
\end{gather*}

\subsection[Class B Reductions mapping $J$ into $-J$]{Class B Reductions mapping $\boldsymbol{J}$ into $\boldsymbol{-J}$}

The class B reductions of the Zakharov--Shabat system change the
sign of $J$, i.e.\ $C^{-1}JC =-J$; therefore we must have also
$\lambda \to -\lambda$.

\begin{remark}\label{rem:-J1}
Note that $\ad_J$ has three eigensubspaces $\mathcal{W}_a$,
$a=0,\pm 1$ corresponding to the eigenvalues 0 and $\pm 2$. The
automorphisms that satisfy $C^{-1}JC =-J$ naturally preserve the
eigensubspace  $\mathcal{W}_0$ but map $\mathcal{W}_{-1}$ onto
$\mathcal{W}_1$ and vice versa. In other words their action on the
root space maps the subset of roots $\Delta_1^+$ onto $\Delta_1^-$
and vice versa: $C\Delta_1^\pm =\Delta_1^\mp $.
\end{remark}

Here we f\/irst consider
\begin{gather*}
    C_5 = S_{e_1-e_2} S_{e_1+e_2} S_{e_3-e_4}  S_{e_3+e_4}C_0.
\end{gather*}
Obviously the product of the the above 4 Weyl ref\/lections will
change the sign of $J$. Its ef\/fect on $Q$ in components reads
\begin{gather*}
q_{ij} = -\epsilon_{ij} q_{ij}^*, \qquad p_{ij} = -\epsilon_{ij}
p_{ij}^*, \qquad \epsilon_{ij} =\epsilon_i \epsilon_j,
\end{gather*}
i.e.\ some of the components of $Q$ become purely imaginary, others
may become real depending on the choice of the signs $\epsilon_j$.

There is no $\bbbz_3$ reduction for the $so(8)$ MMKdV that maps
$J$ into $-J$. So we go directly to the $\bbbz_4$-reduction
generated by $C_6=S_{e_1+e_2} S_{e_2+e_3} S_{e_3+e_4} $  which
maps $J$ into $-J$. The orbits of~$C_3$ are
\begin{gather*}
\mathcal(O)_1^\pm =\{ \pm(e_1+e_2)    \mp (e_2+e_3)  \pm
(e_3+e_4)   \mp (e_1+e_4) \}, \\ \mathcal(O)_2^\pm =\{ \pm
(e_1+e_3)  \mp (e_2+e_4) \}.
\end{gather*}
Again we make use of a convenient basis in $\mathfrak{g}^{(0)}$
\begin{gather*}
\mathcal{E}_\alpha^{(k)} = \sum_{p=0}^3 \ri^{-kp} C_6^{-k} E_\alpha
C_6^k, \qquad \mathcal{F}_\alpha^{(k)} = \sum_{p=0}^3 \ri^{-kp}
C_6^{-k} F_\alpha C_6^k,
\end{gather*}
where $\alpha $ takes values $e_1+e_2$ and $e_1+e_3$. Obviously
\begin{gather}\label{eq:C4m}
C_6^{-1} \mathcal{E}_\alpha^{(k)} C_6 =\ri^k
\mathcal{E}_\alpha^{(k)}, \qquad C_6^{-1} \mathcal{F}_\alpha^{(k)}
C_6 =\ri^k \mathcal{F}_\alpha^{(k)},
\end{gather}
where again $(\mathcal{E}_\alpha^{(0)})^\dag =
\mathcal{F}_\alpha^{(0)}$, $(\mathcal{E}_\alpha^{(k)})^\dag =
\mathcal{F}_\alpha^{(4-k)}$ and $(\mathcal{E}_\alpha^{(k)})^* =
\mathcal{E}_\alpha^{(4-k)}$ for $k=1,2,3$. Then we introduce the
potential
\begin{gather*}
Q(x,t)= \sum_{k=0}^3 \sum_{\alpha } \left( q_\alpha^{(k)}(x,t)
\mathcal{E}_\alpha^{(k)} + p_\alpha^{(k)}(x,t)
\mathcal{F}_\alpha^{(k)} \right).
\end{gather*}
In view of equation (\ref{eq:C4m}) the reduction condition
(\ref{eq:U-V.a}) leads to the following relations between the
coef\/f\/icients
\begin{gather}\label{eq:conZ4m}
p_{\alpha}^{(0)} = (q^{(0)}_{\alpha})^*, \qquad p_{\alpha}^{(k)} =
i^k(q^{(4-k)}_{\alpha})^*, \qquad q_{\alpha}^{(k)} =
i^k (p^{(4-k)}_{\alpha})^*,  \qquad k=1,2,3,
\end{gather}
while the reduction  (\ref{eq:U-V.b}) gives
\begin{gather}
 q_{\alpha}^{(0)} = -(q^{(0)}_{\alpha})^*, \qquad
q_{\alpha}^{(k)} = -i^k(q^{(4-k)}_{\alpha})^*, \qquad
k=1,2, 3, \nonumber\\
 p_{\alpha}^{(0)} = -(p^{(0)}_{\alpha})^*, \qquad
p_{\alpha}^{(k)} = -i^k(p^{(4-k)}_{\alpha})^*, \qquad k=1,2, 3,\label{eq:conZ4m'}
\end{gather}
where $\alpha$ takes values $e_1+e_2$ and $e_1+e_3$. From the
conditions (\ref{eq:conZ4m}) there follows $p_{12}^{(k)} =
q_{12}^{(k)}=p_{13}^{(k)} = q_{13}^{(k)}=0$ for $k=1,3$. In
addition however, it comes up that $\mathcal{E}_{13}^{(0)} =
\mathcal{E}_{13}^{(2)} =\mathcal{F}_{13}^{(0)}=
\mathcal{F}_{13}^{(2)}$. So we are left with only two pairs of
independent functions $p_{12}^{(0)}$, $ q_{12}^{(0)}$ and
$p_{12}^{(2)}$, $ q_{12}^{(2)}$. We provide below slightly more
general formulae for the corresponding Hamiltonian and symplectic
form which are obtained by imposing the constraints
(\ref{eq:conZ4m}) or (\ref{eq:conZ4m'}) and again for simplicity
we skip the upper zeroes in $q_{ij}$ and $p_{ij}$ and replace
$q^{(2)}_{12}$ and $p^{(2)}_{12}$ by $\tilde{q}_{12}$ and
$\tilde{p}_{12}$
\begin{gather*}
H_{\rm MMKdV} = \frac{1}{4}\int_{-\infty}^{\infty} d x\,
\left(\partial^2_{x}q_{12} \partial_{x}p_{12} -\partial_{x} q_{12}
\partial^2_{x}p_{12} +\partial^2_{x}\tilde{q}_{12}
\partial_{x}\tilde{p}_{12} - \partial_{x}\tilde{q}_{12}
\partial^2_{x}\tilde{p}_{12}\right)  \\
\phantom{H_{\rm MMKdV} =}{}  - \frac{3}{32}\int_{-\infty}^{\infty} \Big( \left(
\partial_x (p_{12}^2) + \partial_x (\tilde{q}_{12}^2)\right)
\left( q_{12}^{2} + \tilde{p}_{12}^{2} \right)
 + \partial_x \left(q_{12}\tilde{p}_{12}\right)
p_{12} \tilde{q}_{12}  \\
\phantom{H_{\rm MMKdV} =}{}- \left(\partial_x (q_{12}^2) + \partial_x
(\tilde{p}_{12}^2)\right) \left( p_{12}^{2} + \tilde{q}_{12}^{2}
\right) - \partial_x \left(p_{12}\tilde{q}_{12}\right) q_{12}
\tilde{p}_{12}\Big) d x,
\\
8I_3 =2\int_{-\infty}^{\infty} \left( \partial_{x}q_{12}
\partial_{x}p_{12} + \partial_{x}\tilde{q}_{12}
\partial_{x}\tilde{p}_{12}\right)  d x\\
\phantom{8I_3 =}{}
-\int_{-\infty}^{\infty} \left(  \left(q_{12}p_{12}
+\tilde{q}_{12} \tilde{p}_{12} \right)^2 + (q_{12}\tilde{p}_{12}
+\tilde{q}_{12} p_{12})^2 \right) dx,
\\
\Omega^{(0)} =2\int_{-\infty}^{\infty} d x\, \left(\delta
q_{12}\wedge \delta p_{12} +\delta \tilde{q}_{12}\wedge \delta
\tilde{p}_{12}\right).
\end{gather*}
Now we get two specially coupled mKdV-type equations.

The class B reductions also render all the symplectic forms and
Hamiltonians in the hierarchy real-valued. They allow to render
the corresponding systems of MMKdV equations into ones involving
only real-valued f\/ields. `Half' of the Hamiltonian structures do
not survive these reductions and become degenerate. This holds
true for all symplectic forms $\Omega^{(2m)}$ and integrals of
motion $I_{2m}$ with even indices. However  the other `half' of
the hierarchy with $\Omega^{(2m+1)}$ and integrals of motion
$I_{2m+1}$ remains and provides Hamiltonian properties of the
MMKdV.

Now we get four specially coupled mKdV-type equations given by
\begin{gather*}
\partial_{t} q_{0} +\partial_{x}^{3} q_{0}+ \frac{3}{2}( q_{0}
q_{2} +  p_{0} p_{2})\partial_x p_{2} + \frac{3}{2} (p_{2} q_{2} +
q_{0} p_{0})\partial_x q_{0} =0,\\
\partial_{t} q_{2} +\partial_{x}^{3} q_{2}+ \frac{3}{2} (q_{2}
p_{2}+ q_{0} p_{0})\partial_x q_{2} + \frac{3}{2} (q_{0} q_{2}+
p_{0} p_{2})\partial_x p_{0} =0,\\
\partial_{t} p_{0} +\partial_{x}^{3} p_{0}+ \frac{3}{2} (q_{2}
p_{2}+ q_{0} p_{0})\partial_x p_{0} +\frac{3}{2} (q_{0} q_{2}+
p_{0} p_{2})\partial_x q_{2} =0,\\
\partial_{t} p_{2} +\partial_{x}^{3} p_{2}+ \frac{3}{2} (q_{2}
q_{0}+  p_{0} p_{2})\partial_x q_{0} + \frac{3}{2} (q_{0} p_{0}+
q_{2} p_{2})\partial_x p_{2} =0,
\end{gather*}
where we use for simplicity
\begin{gather*}
q_{12}=q_{0},\qquad \tilde{q}_{12}=q_{2},\qquad
p_{12}=p_{0},\qquad \tilde{p}_{12}=p_{2}
\end{gather*}
and with second reduction (\ref{eq:conZ4m}) $p_{0}=q_{0}^*$ and
$p_{2}=-q_{2}^{*}$ we have
\begin{gather}
\partial_{t} q_{0} +\partial_{x}^{3} q_{0}- \frac{3}{2} (q_{0}
q_{2}- q^{*}_{0} q^{*}_{2})\partial_x q^{*}_{2} - \frac{3}{2} (
q^{*}_{2} q_{2}- q_{0} q^{*}_{0})\partial_x q_{0} =0,\nonumber\\
\partial_{t} q_{2} +\partial_{x}^{3} q_{2} - \frac{3}{2} (q_{2}
q^{*}_{2}- q_{0} q^{*}_{0})\partial_x q_{2} + \frac{3}{2} (q_{0}
q_{2}- q^{*}_{0} q^{*}_{2})\partial_x q^{*}_{0} =0.\label{C80-C6Z2}
\end{gather}
Obviously in the system (\ref{C80-C6Z2}) we can put both
$q_{0}$, $q_{2}$ real with the result
\begin{gather*}
 \partial_{t} q_{0} +\partial_{x}^{3} q_{0}+ \frac{3}{2}(
q_{0}^{2}-q_{2}^2)\partial_x q_{0}=0,\qquad
\partial_{t} q_{2} +\partial_{x}^{3} q_{2} + \frac{3}{2}
(q_{0}^{2}-q_{2}^{2})\partial_x q_{2}=0.
\end{gather*}

The  reduction (\ref{eq:conZ4m'}) means that
$q_\alpha^{(0)}=iq^{\vee}_{0}$, $p_\alpha^{(0)}=ip^{\vee}_{0}$,
$q_\alpha^{(2)}=q^{\vee}_{2}$, $p_\alpha^{(2)}=p^{\vee}_{2}$ with
real valued $p^{\vee}_{i}$, $q^{\vee}_{i}$, $i=0,2$. Thus we get
\begin{gather*}
 \partial_{t} q^{\vee}_{0} +\partial_{x}^{3} q^{\vee}_{0}+
\frac{3}{2} (q^{\vee}_{0} q^{\vee}_{2}+ p^{\vee}_{0} p^{\vee}_{2})
\partial_x p^{\vee}_{2} + \frac{3}{2} (p^{\vee}_{2}
q^{\vee}_{2}- q^{\vee}_{0} p^{\vee}_{0})\partial_x q^{\vee}_{0}
=0,\\
\partial_{t} q^{\vee}_{2} +\partial_{x}^{3} q^{\vee}_{2}+ \frac{3}{2} (q^{\vee}_{2}
p^{\vee}_{2}- q^{\vee}_{0} p^{\vee}_{0})\partial_x q^{\vee}_{2} -
\frac{3}{2} (q^{\vee}_{0} q^{\vee}_{2}+ p^{\vee}_{0}
p^{\vee}_{2})\partial_x p^{\vee}_{0} =0,\\
\partial_{t} p^{\vee}_{0} +\partial_{x}^{3} p^{\vee}_{0}+ \frac{3}{2} (q^{\vee}_{2}
p^{\vee}_{2}-q^{\vee}_{0} p^{\vee}_{0})\partial_x p^{\vee}_{0} +
\frac{3}{2} (q^{\vee}_{0} q^{\vee}_{2}+ p^{\vee}_{0}
p^{\vee}_{2})\partial_x q^{\vee}_{2} =0,\\
\partial_{t} p^{\vee}_{2} +\partial_{x}^{3} p^{\vee}_{2}- \frac{3}{2} (q^{\vee}_{2}
q^{\vee}_{0}+ p^{\vee}_{0} p^{\vee}_{2})\partial_x q^{\vee}_{0}  -
\frac{3}{2} (q^{\vee}_{0} p^{\vee}_{0}- q^{\vee}_{2}
p^{\vee}_{2})\partial_x p^{\vee}_{2} =0.
\end{gather*}

The class B reductions also render all the symplectic forms and
Hamiltonians in the hierarchy real-valued. They allow to render
the corresponding systems of MMKdV equations into ones involving
only real-valued f\/ields. `Half' of the Hamiltonian structures do
not survive these reductions and become degenerate. This holds
true for all symplectic forms $\Omega^{(2m)}$ and integrals of
motion $I_{2m}$ with even indices. However  the other `half' of
the hierarchy with $\Omega^{(2m+1)}$ and integrals of motion
$I_{2m+1}$ remains and provides Hamiltonian properties of the
MMKdV.

\subsection[Effects of class A reductions on the scattering data]{Ef\/fects of class A reductions on the scattering data}

The ref\/lection coef\/f\/icients $\rho ^\pm(\lambda ) $ and
$\tau^\pm(\lambda ) $ are def\/ined only on the real $\lambda
$-axis, while the diagonal blocks $\a^\pm(\lambda ) $ and
$\c^\pm(\lambda ) $ (or, equivalently, $D^\pm(\lambda ) $) allow
analytic extensions for $\lambda \in \bbbc_\pm $. From the
equations (\ref{eq:6.2}) there follows that
\begin{gather}\label{eq:2x2-1}
\a^+(\lambda )\c^-(\lambda ) = (\openone +\rho ^-\rho ^+(\lambda
))^{-1}, \qquad \a^-(\lambda )\c^+(\lambda) =(\openone +\rho^+\rho
^-(\lambda
))^{-1},\\
\label{eq:2x2-2} \c^-(\lambda )\a^+(\lambda ) = (\openone
+\tau^+\tau ^-(\lambda ))^{-1}, \qquad \c^+(\lambda )\a^-(\lambda
) = (\openone +\tau ^-\tau ^+(\lambda ))^{-1}.
\end{gather}

Given $\mathcal{T}_1 $ (resp.~$\mathcal{T}_2 $) we determine the
right hand sides of  (\ref{eq:2x2-1}) (resp.~(\ref{eq:2x2-2})) for
$\lambda \in \bbbr $. Combined with the facts about the limits
\begin{gather}\label{eq:2x2-3}
\lim_{\lambda \to\infty } \a^+(\lambda ) =\openone , \qquad
\lim_{\lambda \to\infty } \c^-(\lambda ) =\openone ,
\qquad
\lim_{\lambda \to\infty } \a^-(\lambda ) =\openone , \qquad
\lim_{\lambda \to\infty } \c^+(\lambda ) =\openone ,
\end{gather}
each of the relations (\ref{eq:2x2-1}), (\ref{eq:2x2-2}) can be
viewed as a RHP with canonical normalization. Such RHP can be
solved explicitly in the one-component case (provided we know the
locations of their zeroes) by using the Plemelj--Sokhotsky formulae~\cite{Gakhov}. These zeroes are in fact the discrete eigenvalues
of~$L $. One possibility to make use of these facts is to take log
of the determinants of both sides of (\ref{eq:2x2-1}) getting
\begin{gather*}
A^+(\lambda )+C^-(\lambda ) = - \ln \det (\openone +\rho ^-\rho
^+(\lambda ), \qquad \lambda \in \bbbr,
\end{gather*}
where
\begin{gather*}
A^\pm (\lambda ) = \ln \det \a^\pm(\lambda ), \qquad C^\pm
(\lambda ) = \ln \det \c^\pm(\lambda ).
\end{gather*}
Then Plemelj--Sokhotsky formulae  allows us to recover
$A^\pm(\lambda ) $ and $C^\pm(\lambda ) $
\begin{gather}\label{eq:2x2-6}
\mathcal{A}(\lambda ) = {\ri \over 2\pi } \int_{-\infty }^{\infty
} { \rd\mu\over \mu -\lambda } \ln \det (\openone +\rho ^-\rho
^+(\mu )) + \sum_{j=1}^{N} \ln {\lambda -\lambda _j^+ \over
\lambda -\lambda _j^- },
\end{gather}
where $\mathcal{A}(\lambda )=A^+(\lambda ) $ for $\lambda \in
\bbbc_+ $ and $\mathcal{A}(\lambda )=-C^-(\lambda ) $ for $\lambda
\in \bbbc_-$. In deriving (\ref{eq:2x2-6}) we have also assumed
that $\lambda _j^\pm $ are simple zeroes of $A^\pm(\lambda ) $ and
$C^\pm(\lambda ) $.

Let us consider a reduction condition (\ref{eq:U-V.a}) with $C_1$
from the Cartan subgroup: $C_1=\diag(B_+,B_-) $ where the diagonal
matrices $B_\pm$ are such that $B_\pm^2 =\openone $. Then we get
the following constraints on the sets $\mathcal{T}_{1,2} $
\begin{gather*}
\rho ^-(\lambda )=(B_-\rho ^+(\lambda )B_+)^\dag, \qquad \rho_j
^-=(B_-\rho_j ^+B_+)^\dag , \qquad \lambda _j^-=(\lambda _j^+)^*,
\\ 
\tau^-(\lambda )=(B_+\tau ^+(\lambda )B_-)^\dag,
\qquad \tau_j ^-=(B_+\tau_j ^+B_-)^\dag , \qquad \lambda_j^-
=(\lambda _j^+)^*,
\end{gather*}
where $ j=1,\dots, N$. For more details see Subsections~\ref{ssec:2x} and~\ref{sec4.4}.

\begin{remark}\label{rem:3}
For certain  reductions  such as, e.g.\ $Q=-Q^\dag $ the
generalized Zakharov--Shabat system $L(\lambda )\psi =0 $ can be
written down as an eigenvalue problem $\mathcal{L}\psi =\lambda
\psi (x,\lambda ) $ where $\mathcal{L} $ is a self-adjoint
operator. The continuous spectrum of $\mathcal{L} $ f\/ills up the
whole real $\lambda $-axis thus `leaving no space' for discrete
eigenvalues. Such Lax operators have no discrete spectrum and the
corresponding MNLS or MMKdV equations do not have soliton
solutions.

\end{remark}

From the general theory of RHP \cite{Gakhov} one may conclude that
(\ref{eq:2x2-1}), (\ref{eq:2x2-2}) allow unique solutions provided
the number and types of the zeroes $\lambda _j^\pm $ are properly
chosen. Thus we can outline a procedure which allows one to
reconstruct not only $T(\lambda ) $ and $\hat{T}(\lambda ) $ and
the corresponding potential $Q(x) $ from each of the sets
$\mathcal{T}_i $, $i=1,2 $:

\begin{description}\itemsep=0pt

\item [i)] Given $\mathcal{T}_2 $ (resp.~$\mathcal{T}_1 $) solve
the RHP (\ref{eq:2x2-1}) (resp.~(\ref{eq:2x2-2})) and construct
$\a^\pm(\lambda ) $ and $\c^\pm(\lambda ) $ for $\lambda
\in\bbbc_\pm $.

\item [ii)] Given $\mathcal{T}_1 $ we determine $\b^\pm(\lambda )
$ and $\d^\pm(\lambda ) $ as
\begin{gather*}
\b^\pm(\lambda ) =\rho ^\pm(\lambda )\a^\pm(\lambda ) , \qquad
\d^\pm(\lambda ) =\c^\pm(\lambda )\rho ^\pm(\lambda ) ,
\end{gather*}
or if $\mathcal{T}_2 $ is known then
\begin{gather*}
\b^\pm(\lambda ) =\a^\pm(\lambda )\tau ^\pm(\lambda ) ,\qquad
\d^\pm(\lambda ) =\tau ^\pm(\lambda )\c^\pm(\lambda ) .
\end{gather*}

\item [iii)] The potential $Q(x) $ can be recovered from
$\mathcal{T}_1 $ by solving the RHP (\ref{eq:12.4}) and using equation~(\ref{eq:Q'}).

\end{description}

Another method for reconstructing $Q(x) $ from $\mathcal{T}_j $
uses the interpretation of the ISM as gene\-ra\-lized Fourier
transform, see \cite{AKNS,KN79,G}.

Let in this subsection all automorphisms $C_i$ are of class A.
Therefore acting on the root space they preserve the vector
$\sum\limits_{k=1}^r e_k$ which is dual to $J$, and as a consequence, the
corresponding Weyl group elements map the subset of roots
$\Delta_1^+$ onto itself.

\begin{remark}\label{rem:n4}
An important consequence of this is that $C_i$ will map
block-upper-triangular (resp. block-lower-triangular) matrices
like in equation~(\ref{eq:6.4}) into matrices with the same block
structure.  The block-diagonal matrices will be mapped again into
block-diagonal ones.
\end{remark}

From the reduction conditions (\ref{eq:U-V.a})--(\ref{eq:U-V.d})
one gets, in the limit $x\to\infty$ that
\begin{gather}
 \mbox{a)}\quad C_1((\kappa_1(\lambda)J)^\dag) = \lambda J, \qquad
\mbox{b)}\quad C_2((\kappa_2(\lambda)J)^T) = -\lambda J,\nonumber \\
 \mbox{c)}\quad C_3((\kappa_3(\lambda)J)^*) = \lambda J,
\qquad  \mbox{d)}\quad C_4((\kappa_4(\lambda)J)) = \lambda J.\label{eq:63.1}
\end{gather}
Using equation (\ref{eq:63.1}) and $C_i(J)=J$ one f\/inds that:
\begin{gather*}
\mbox{a)}\quad \kappa_1(\lambda) =\lambda^* , \qquad \mbox{b)}\quad
\kappa_2(\lambda) =-\lambda, \qquad  \mbox{c)}\quad
\kappa_3(\lambda) = -\lambda^*,  \qquad  \mbox{d)}\quad
\kappa_4(\lambda) = \lambda.
\end{gather*}

It remains to take into account that the reductions
(\ref{eq:U-V.a})--(\ref{eq:U-V.d}) for the potentials of $L$ lead
to the following constraints on the scattering matrix
$T(\lambda)$
\begin{alignat*}{5}
 & \mbox{a)}\quad && C_1(T^\dag(\lambda^*)) = \hat{T}(\lambda), \qquad&&
\mbox{b)}\quad&& C_2(T^T(-\lambda)) = \hat{T}(\lambda ),& \\
 & \mbox{c)}\quad && C_3((T^*(-\lambda^*)) = T(\lambda), \qquad &&
\mbox{d)}\quad && C_4(T(\lambda)) = T(\lambda ).&
\end{alignat*}

These results along with Remark~\ref{rem:n4} lead to the following
results for the generalized Gauss factors of $T(\lambda)$
\begin{alignat*}{4}\label{eq:63.3}
& \mbox{a)}\quad && C_1(\S^{+,\dag}(\lambda^*)) = \hat{\S^-}(\lambda),
\qquad && C_1(\T^{-,\dag}(\lambda^*)) = \hat{\T^+}(\lambda), &  \\
& \mbox{b)}\quad && C_2(\S^{+,T}(-\lambda)) = \hat{\S^-}(\lambda),
\qquad && C_2(\T^{-,T}(-\lambda)) = \hat{\T^+}(\lambda), &  \\
& \mbox{c)}\quad && C_3(\S^{\pm,*}(-\lambda^*)) = \S^\pm(\lambda),
\qquad && C_3(\T^{\pm,*}(-\lambda^*)) = \T^\pm(\lambda), &  \\
&\mbox{d)}\quad && C_4(\S^{\pm}(\lambda)) = \S^\pm(\lambda), \qquad &&
C_4(\T^{\pm}(\lambda)) = \T^\pm(\lambda)&
\end{alignat*}
and
\begin{alignat*}{5}
& \mbox{a)}\quad && C_1(D^{+,\dag}(\lambda^*)) = \hat{D^-}(\lambda),
\qquad && \mbox{b)}\quad && C_2(D^{+,T}(-\lambda)) = \hat{D^-}(\lambda), & \\
& \mbox{c)}\quad && C_3(D^{\pm,*}(-\lambda)) = D^\pm(\lambda),
\qquad &&  \mbox{d)}\quad && C_4(D^{\pm}(\lambda)) = D^\pm(\lambda).&
\end{alignat*}

\subsection[Effects of class B reductions on the scattering data]{Ef\/fects of class B reductions on the scattering data}

In this subsection all automorphisms $C_i$ are of class B.
Therefore acting on the root space they map the vector
$\sum\limits_{k=1}^r e_k$ dual to $J$ into $-\sum\limits_{k=1}^r e_k$. As a
consequence, the corresponding Weyl group elements map the subset
of roots $\Delta_1^+$ onto $\Delta_1^- \equiv -\Delta_1^+$.

\begin{remark}\label{rem:n5}
An important consequence of this is that $C_i$ will map
block-upper-triangular into block-lower-triangular matrices like
in equation~(\ref{eq:6.4}) and vice versa. The block-diagonal matrices
will be mapped again into block-diagonal ones.
\end{remark}

Now equation (\ref{eq:63.1}) with $C_i(J)=-J$ leads to
\begin{gather*}
 \mbox{a)}\quad \kappa_1(\lambda) =-\lambda^* , \qquad
\mbox{b)}\quad \kappa_2(\lambda) =\lambda , \qquad
 \mbox{c)}\quad \kappa_3(\lambda) = \lambda^* , \qquad
\mbox{d)}\quad \kappa_4(\lambda) = -\lambda.
\end{gather*}

The reductions (\ref{eq:U-V.a})--(\ref{eq:U-V.d}) for the
potentials of $L$ lead to the following constraints on the
scattering matrix $T(\lambda)$:
\begin{alignat*}{5}
& \mbox{a)}\quad && C_1(T^\dag(-\lambda^*)) = \hat{T}(\lambda), \qquad &&
\mbox{b)}\quad && C_2(T^T(\lambda)) = \hat{T}(\lambda ), & \\
& \mbox{c)}\quad && C_3((T^*(\lambda^*)) = T(\lambda), \qquad &&
\mbox{d)}\quad && C_4(T(-\lambda)) = T(\lambda ).&
\end{alignat*}

Then along with Remark \ref{rem:n5} we f\/ind the following results
for the generalized Gauss factors of~$T(\lambda)$
\begin{alignat*}{4}
& \mbox{a)}\quad && C_1(\S^{\pm,\dag}(-\lambda^*)) =
\hat{\S^\pm}(\lambda),
\qquad && C_1(\T^{\pm,\dag}(-\lambda^*)) = \hat{\T^\pm}(\lambda), &  \\
& \mbox{b)}\quad && C_2(\S^{\pm,T}(\lambda)) = \hat{\S^\pm}(\lambda),
\qquad && C_2(\T^{\pm,T}(\lambda)) = \hat{\T^\pm}(\lambda), &  \\
& \mbox{c)}\quad && C_3(\S^{+,*}(\lambda^*)) = \S^-(\lambda),
\qquad && C_3(\T^{-,*}(\lambda^*)) = \T^+(\lambda), &  \\
&\mbox{d)}\quad && C_4(\S^{+}(-\lambda)) = \S^-(\lambda), \qquad &&
C_4(\T^{-}(-\lambda)) = \T^+(\lambda) &
\end{alignat*}
and
\begin{alignat*}{5} 
& \mbox{a)}\quad && C_1(D^{\pm,\dag}(-\lambda^*)) =
\hat{D^\pm}(\lambda),
\qquad && \mbox{b)}\quad && C_2(D^{\pm,T}(\lambda)) = \hat{D^\pm}(\lambda), & \\
& \mbox{c)} \quad && C_3(D^{+,*}(\lambda^*)) = D^-(\lambda),
\qquad &&  \mbox{d)}\quad && C_4(D^{+}(-\lambda)) = D^-(\lambda).&
\end{alignat*}

\section[The classical $r$-matrix and the NLEE of MMKdV type]{The classical $\boldsymbol{r}$-matrix and the NLEE of MMKdV type}
\label{ssec:5.3}

     One of the def\/initions of the classical $r$-matrix is based on
the Lax representation for the corresponding NLEE. We will start
from this def\/inition, but before to state it will introduce the
following notation
\begin{gather*}
\Big\{U(x,\lambda ) \otimescomma U(y,\mu )\Big\}  ,
\end{gather*}
which is an abbreviated record for the Poisson bracket between all
matrix elements of $U(x,\lambda )$ and $U(y,\mu )$
\begin{gather*}
\Big\{ U(x,\lambda ) \otimescomma U(y,\mu )\Big\}_{ik,lm} =
\left\{U_{ik}  (x,\lambda ), U_{lm}  (y,\mu )\right\} .
\end{gather*}
In particular, if $U(x,\lambda )$ is of the form
\begin{gather*}
U(x,\lambda ) = Q(x,t) - \lambda J  ,\qquad Q(x,t) = \sum_{i<r}^{}
(q_{ir}E_{ir}  + p_{ri} E_{ri}) ,
\end{gather*}
and the matrix elements of $Q(x,t)$ satisfy canonical Poisson
brackets
\begin{gather*}
\{q_{ks},p_{ri}\}=\ri\delta_{ik}\delta_{rs}\delta(x-y),
\end{gather*}
then
\begin{gather*}
\Big\{U(x,\lambda ) \otimescomma U(y,\mu )\Big\} = \ri
\sum_{i<r}^{}( E_{ir}\otimes E_{ri} - E_{ri}\otimes E_{ir}) \delta
(x-y) .
\end{gather*}

The classical $r$-matrix can be def\/ined through the relation~\cite{FaTa}
\begin{gather}\label{eq:L5.5}
\Big\{U(x,\lambda ) \otimescomma  U(y,\mu )\Big\} =
\ri\left[r(\lambda -\mu ),U(x,\lambda )\otimes \openone  +
\openone \otimes U(y,\mu )\right] \delta (x-y),
\end{gather}
which can be understood as a system of $N^2 $ equation for the
$N^2 $ matrix elements of \mbox{$r(\lambda -\mu )$}. However, these
relations must hold identically with respect to $\lambda $ and
$\mu $, i.e., (\ref{eq:L5.5}) is an overdetermined system of
algebraic equations for the matrix elements of $r$. It is far from
obvious whether such $r(\lambda -\mu )$ exists, still less obvious
is that it depends only on the dif\/ference $\lambda -\mu $. In
other words far from any choice for $U(x,\lambda )$ and for the
Poisson brackets between its matrix elements allow $r$-matrix
description. Our system  (\ref{eq:L5.5}) allows an  $r$-matrix
given by
\begin{gather}\label{eq:L5.6}
r(\lambda -\mu ) = - {1\over 2 } { P \over \lambda  - \mu  } ,
\end{gather}
where $P$ is a constant $N^2\times N^2$ matrix
\begin{gather*}
P = \sum_{\alpha\in \Delta^+}^{N} (E_{\alpha}\otimes E_{-\alpha} +
E_{-\alpha}\otimes E_{\alpha}) + \sum_{k=1}^r h_k\otimes h_k.
\end{gather*}

     The matrix $P$ possesses the following special properties
\begin{gather*}
[P, X \otimes  \openone + \openone \otimes  X] = 0 \qquad \forall\,
X\in \mathfrak{g}.
\end{gather*}
By using these properties of $P$ we are getting
\begin{gather}\label{eq:L5.9}
\left[ P, Q(x)\otimes \openone  + \openone \otimes Q(x)\right] = 0
,
\end{gather}
i.e., the r.h.s.\ of (\ref{eq:L5.5}) does not contain $Q(x,t)$.
Besides:
\begin{gather}
  \left[P, \lambda J\otimes \openone  + \mu \openone \otimes
J\right] =
(\lambda  - \mu ) \left[P, J \otimes \openone \right] \nonumber\\
\qquad  {}= - 2(\lambda  - \mu )\left( \sum_{\alpha\in\Delta_1^+}^{}(
E_{\alpha}\otimes E_{-\alpha} - E_{-\alpha}\otimes E_{\alpha}) \right) ,\label{eq:L5.10}
\end{gather}
where we used the commutation relations between the elements of
the Cartan--Weyl basis. The comparison between (\ref{eq:L5.9}),
(\ref{eq:L5.10}) and (\ref{eq:L5.5}) leads us to the result, that
$r(\lambda -\mu )$ (\ref{eq:L5.6}) indeed satisf\/ies the def\/inition
(\ref{eq:L5.5}).

\begin{remark}
It is easy to prove that equation (\ref{eq:L5.5}) is invariant
under the reduction conditions~(\ref{eq:U-V.a}) and
(\ref{eq:U-V.b}) of class A, so the  corresponding reduced
equations have the same $r$-matrix. However (\ref{eq:L5.5}) is not
invariant under the class B reductions and the corresponding
reduced NLEE do not allow classical $r$-matrix def\/ined through equation~(\ref{eq:L5.5}). \end{remark}

Let us now show, that the classical $r$-matrix is a very
ef\/fective tool for calculating the Poisson brackets between the
matrix elements of $T (\lambda )$. It will be more convenient here
to consider periodic boundary conditions on the interval $[-L,L]
$, i.e.\ $Q(x-L)=Q(x+L) $ and to use the fundamental solution
$T(x,y,\lambda ) $ def\/ined by
\begin{gather*}
\ri {\rd T(x,y,\lambda )  \over \rd x } + U(x,\lambda
)T(x,y,\lambda ) =0, \qquad T(x,x,\lambda )=\openone .
\end{gather*}

Skipping the details we just formulate the following relation for
the Poisson brackets between the matrix elements of $T(x,y,\lambda
) $ \cite{FaTa}
\begin{gather}\label{eq:L5.12}
\Big\{ T(x,y,\lambda ) \otimescomma  T(x,y,\mu )\Big\} = \left[
r(\lambda -\mu ), T(x,y,\lambda )\otimes T(x,y,\mu )\right].
\end{gather}

The corresponding monodromy matrix $T_L(\lambda ) $ describes the
transition from $-L $ to $L $ and $T_L(\lambda ) =T(-L,L,\lambda
)$. The Poisson brackets between the matrix elements of
$T_L(\lambda )$ follow directly from equation~(\ref{eq:L5.12}) and are
given by
\begin{gather*}
\Big\{T_L (\lambda ) \otimescomma T_L(\mu )\Big\} =
\left[r(\lambda -\mu ), T_L(\lambda ) \otimes T_L (\mu )\right]  .
\end{gather*}

An elementary consequence of this result is the involutivity of
the integrals of motion $I_{L,k}$ from the principal series which
are from the expansions of
\begin{gather}
\ln \det \a_L^+(\lambda ) = \sum_{k=1}^{\infty } I_{L,k}\lambda
^{-k}, \qquad  -\ln \det \c_L^-(\lambda ) = \sum_{k=1}^{\infty }
I_{L,k}\lambda
^{-k}, \nonumber\\
\ln \det \c_L^+(\lambda ) = \sum_{k=1}^{\infty } J_{L,k}\lambda
^{-k}, \qquad  -\ln \det \a_L^-(\lambda ) = \sum_{k=1}^{\infty }
J_{L,k}\lambda ^{-k}.\label{eq:ln-det}
\end{gather}
An important property of the integrals $I_{L,k} $ and $J_{L,k} $
is their locality, i.e.\ their densities depend only on $Q $ and
its $x $-derivatives.

The simplest consequence of the  relation (\ref{eq:L5.12}) is the
involutivity of $I_{L,k} $, $J_{L,k} $. Indeed, taking the trace
of both sides of (\ref{eq:L5.12}) shows that $\{ \tr T_L(\lambda
),\tr T_L(\mu )\}=0 $.  We can also multiply both sides of
(\ref{eq:L5.12}) by $C\otimes C$ and then take the trace using equation
(\ref{eq:L5.9}); this proves
\begin{gather*}
\left\{\tr T_L (\lambda )C, \tr T_L (\mu )C\right\} = 0 .
\end{gather*}
In particular, for $C = \openone +J$  and $C = \openone -J$  we
get the involutivity of
\begin{gather*}
\left\{ \tr \a^+_L(\lambda ),\tr \a^+_L(\mu )\right\} =0, \qquad
\left\{\tr \a^-_L(\lambda ),\tr \a^-_L(\mu )\right\} = 0 ,\\
\left\{ \tr \c^+_L(\lambda ),\tr \c^+_L(\mu )\right\} =0, \qquad
\left\{\tr \c^-_L(\lambda ),\tr \c^-_L(\mu )\right\} = 0.
\end{gather*}

Equation~(\ref{eq:L5.12}) was derived for the typical representation
$V^{(1)} $ of $\mathfrak{G}\simeq SO(2r) $, but it holds true also
for any other f\/inite-dimensional representation of $\mathfrak{G}
$. Let us denote by $V^{(k)}\simeq \wedge^{k}V^{(1)} $ the $k $-th
fundamental representation of $\mathfrak{G} $; then the element
$T_L(\lambda ) $ will be represented in $V^{(k)} $ by~$\wedge^k
T_L(\lambda ) $~-- the $k $-th wedge power of $T_L(\lambda ) $,
see \cite{Helg}. In particular, if we consider equation~(\ref{eq:L5.12}) in the representation $V^{(n)} $ and sandwich it
between the highest and lowest weight vectors in $V^{(n)} $ we
get \cite{PLA126}
\begin{gather}\label{eq:det-a}
\{ \det \a^+_L(\lambda ), \det \a^+_L(\mu )\} =0, \qquad \{ \det
\c^-_L(\lambda ), \det \c^-_L(\mu )\} =0.
\end{gather}

Likewise considering (\ref{eq:L5.12}) in the representation
$V^{(m)} $ and sandwich it between the highest and lowest weight
vectors in $V^{(m)} $ we get
\begin{gather}\label{eq:det-am}
\{ \det \a^-_L(\lambda ), \det \a^-_L(\mu )\} =0, \qquad \{ \det
\c^+_L(\lambda ), \det \c^+_L(\mu )\} =0.
\end{gather}
Since equations (\ref{eq:det-a}) and (\ref{eq:det-am}) hold true for
all values of $\lambda  $ and $\mu  $ we can insert into them the
expansions (\ref{eq:ln-det}) with the result
\begin{gather*}
\{ I_{L,k}, I_{L,p}\} =0,\qquad \{ J_{L,k}, J_{L,p}\} =0,\qquad
k,p= 1,2,\dots.
\end{gather*}

Somewhat more general analysis along this lines allows one to see
that only the eigenvalues of $\a_L^\pm(\lambda ) $ and
$\c_L^\pm(\lambda ) $ produce integrals of motion in involution.

Taking the limit $L \to \infty $ we are able to transfer these
results also for the case of potentials with zero boundary
conditions. Indeed, let us multiply (\ref{eq:L5.12}) by
$E(y,\lambda ) \otimes E(y,\mu )$ on the right and by $E^{-1}
(x,\lambda ) \otimes E^{-1} (x,\mu )$  on the left, where
$E(x,\lambda)=\exp (-\ri\lambda Jx)$ and take the limit for $x \to
\infty $, $y \to -\infty $. Since
\begin{gather*}
\lim_{x\to\pm \infty } {\re^{\ri x(\lambda -\mu )} \over \lambda
- \mu } =\pm \ri \pi  \delta (\lambda  - \mu ),
\end{gather*}
we get
\begin{gather*}
\Big\{T (\lambda ) \otimescomma T (\mu )\Big\} = r_+ (\lambda
- \mu )T (\lambda ) \otimes T (\mu ) -
T (\lambda ) \otimes T (\mu ) r_- (\lambda -\mu ) , \\
 r_\pm (\lambda -\mu ) = - {1\over 2(\lambda  - \mu ) } \left( \sum_{k=1}^r h_k\otimes h_k +
 \sum_{\alpha\in\Delta_0^+}
 (E_{\alpha} \otimes E_{-\alpha}+E_{-\alpha} \otimes E_{\alpha}) \right)\!
 \pm \ri\pi \delta (\lambda -\mu )\Pi_{0J},
\end{gather*}
where $\Pi_{0J}$ is def\/ined by
\begin{gather*}
\Pi_{0J}=\sum_{\alpha\in\Delta_1^+}(E_{\alpha}\otimes E_{-\alpha}-
E_{-\alpha}\otimes E_{\alpha} ).
\end{gather*}
Analogously we prove that
i) the integrals $I_k=\lim\limits_{L\to\infty }I_{L,k} $ and
$J_p=\lim\limits_{L\to\infty }J_{L,p} $ are in involution, i.e.
\[ \{I_k, I_p\} =  \{I_k, J_p\} =  \{J_k, J_p\} = 0, \]
for all positive values of $k $ and $p $; ii) only the eigenvalues
of $\a^\pm(\lambda ) $ and $\c^\pm(\lambda ) $ produce integrals
of motion in involution.

\section{Conclusions}\label{sec:5}
We showed that the interpretation of the ISM as a generalized
Fourier transform holds true also for the generalized
Zakharov--Shabat systems related to the symmetric spaces {\bf DIII}. The expansions over the `squared solutions' are natural
tool to derive the fundamental properties not only of the MNLS
type equations, but also of the NLEE with generic dispersion laws,
in particular MMKdV equations. Some of these equations, besides
the intriguing properties as dynamical systems allowing for
boomerons, trappons etc., may also have interesting physical
applications.

Another interesting area for further investigations is to study
and classify the reductions of these NLEE. For results along this
line for the MNLS equations see the reports \cite{Varna04} and
\cite{manev04}; reductions of other types of NLEE have been
considered in \cite{G-cm,vgn,vgrn,LoMi1}.

One can also treat generalized Zakharov--Shabat systems related to
other symmetric spaces. The expansions over the `squared
solutions' can be closely related to the graded Lie algebras, and
to the reduction group and provide an ef\/fective tool to derive and
analyze new soliton equations. For more details and further
reading see \cite{DrSok,G-cm}.

In conclusion, we have considered the reduced multicomponent MKdV
equations associated with the {\bf DIII}-type symmetric
spaces using A.V.~Mikhailov reduction group. Several examples of
such nontrivial reductions leading to new MMKdV systems related to
the $so(8)$ Lie algebra are given. In particular we provide
examples with reduction groups isomorphic to $\bbbz_{2}$,
$\bbbz_{3}$, $\bbbz_{4}$ and derive their ef\/fects on the
scattering matrix, the minimal sets of scattering data and on the
hierarchy of Hamiltonian structures. These results can be
generalized also for other types of symmetric spaces.

\subsection*{Acknowledgements}

It is our pleasure to thank Professor Gaetano Vilasi for useful
discussions. One of us (VSG) thanks the Physics Department of the
University of Salerno for the kind hospitality and the Italian
Istituto Nazionale di Fisica Nucleare for f\/inancial support. We
also thank the Bulgarian Science Foundation for partial support
through  contract No.~F-1410. One of us (VSG) is grateful to the
organizers of the Kyiv conference for their hospitality and
acknowledges f\/inancial support by an CEI grant.

\pdfbookmark[1]{References}{ref}
\LastPageEnding

\end{document}